\documentclass[twocolumn]{aastex631}

\newcommand{\teff}{$T_{\mathrm{eff}}$}
\newcommand{\logg}{$\log g$}

\newcommand{\kep}{{\it Kepler}}

\newcommand{\gaia}{{\it Gaia}}

\newcommand{\prot}{$P_{\mathrm{rot}}$}
\newcommand{\vect}[1]{\boldsymbol{#1}}

\usepackage{hyperref}
\usepackage{amsmath}

\makeatletter
\newcommand{\labeltext}[2]{%
  \@bsphack
  \csname phantomsection\endcsname 
  \def\@currentlabel{#1}{\label{#2}}%
  \@esphack
}

\shorttitle{[Y/Mg] Clock for FGK Stars}
\shortauthors{Berger et al.}

\received{6/3/2022}
\revised{8/3/2022}
\accepted{8/3/2022}
\submitjournal{ApJ}

\begin{document}

\title{Is [Y/Mg] a Reliable Age Diagnostic for FGK Stars?}

\correspondingauthor{Travis Berger}
\email{travis.a.berger@nasa.gov}

\author[0000-0002-2580-3614]{Travis A. Berger}
\altaffiliation{NASA Postdoctoral Program Fellow}
\affiliation{Exoplanets and Stellar Astrophysics Laboratory, Code 667, NASA Goddard Space Flight Center, Greenbelt, MD, 20771, USA}
\affiliation{Institute for Astronomy, University of Hawai`i, 2680 Woodlawn Drive, Honolulu, HI 96822, USA}

\author[0000-0002-4284-8638]{Jennifer L. van Saders}
\affiliation{Institute for Astronomy, University of Hawai`i, 2680 Woodlawn Drive, Honolulu, HI 96822, USA}

\author[0000-0001-8832-4488]{Daniel Huber}
\affiliation{Institute for Astronomy, University of Hawai`i, 2680 Woodlawn Drive, Honolulu, HI 96822, USA}

\author[0000-0002-5258-6846]{Eric Gaidos}
\affiliation{Department of Earth Sciences, University of Hawai`i at M\={a}noa, Honolulu, HI 96822, USA}

\author[0000-0001-5347-7062]{Joshua E. Schlieder}
\affiliation{Exoplanets and Stellar Astrophysics Laboratory, Code 667, NASA Goddard Space Flight Center, Greenbelt, MD, 20771, USA}

\author[0000-0002-9879-3904]{Zachary R. Claytor}
\affiliation{Institute for Astronomy, University of Hawai`i, 2680 Woodlawn Drive, Honolulu, HI 96822, USA}

\begin{abstract}
Current spectroscopic surveys are producing large catalogs of chemical abundances for stars of all types. The yttrium to magnesium ratio, [Y/Mg], has emerged as a candidate age indicator for solar twins in the local stellar neighborhood. However, it is unclear whether it is a viable age diagnostic for more diverse stellar types, so we investigate [Y/Mg] as an age indicator for the FGK-type planet host stars observed by $Kepler$. We find that the [Y/Mg] ``Clock'' is most precise for solar twins, with a [Y/Mg]/Age slope of $m$ = --0.0370 $\pm$ 0.0071 dex/Gyr and $\sigma_{\mathrm{Age}}$ = 2.6 Gyr. We attribute the lower precision compared to literature results to non-solar twins contaminating our solar twin sample and recommend a 1.5 Gyr systematic uncertainty for stellar ages derived with any [Y/Mg]-Age relation. We also analyzed the [Y/Mg] Clock as a function of \teff, \logg, and metallicity individually and find no strong trends, but compute statistically significant [Y/Mg]-Age relations for subsamples defined by ranges in \teff, \logg, and metallicity. Finally, we compare [Y/Mg] and rotation ages and find statistically similar trends as for isochrone ages, although we find that rotation ages perform better for GK dwarfs while isochrones perform better for FG subgiants. We conclude that the [Y/Mg] Clock is most precise for solar twins and analogs but is also a useful age diagnostic for FGK stars.
\end{abstract}

\section{Introduction}

Stellar ages are invaluable for interpreting the sequence of events in the universe. However, stellar ages are typically very difficult to estimate for field stars \citep{Soderblom2010}. Field star ages have been determined through isochrone comparison \citep[or placement in a color-magnitude diagram,][]{Edvardsson1993,GCS1,GCS3,Morton2016,Johnson2017}, now more powerful with \gaia\ parallaxes \citep{Fulton2018,Berger2018c,Berger2020a,Berger2020b}, to gyrochronology \citep{Barnes2007,Mamajek2008,vansaders16,Angus2018,Curtis2019} to asteroseismology \citep{otifloranes05,mazumdar05,kjeldsen08b,Aguirre2015,Creevey2017,Pinsonneault2018}, and to galactic kinematics \citep{Makarov2007,Fernandez2008,Angus2020,Lu2021}.

Stellar ages can also be inferred from chemical abundances, such as lithium abundances \citep{Skumanich1972,Soderblom1993,Sestito2005,Mentuch2008,Boesgaard2016,Berger2018,Deepak2019,Gaidos2020,Magrini2021}, which are expected to evolve either during the lifetime of a star or over the lifetimes of many stars as they enrich the interstellar medium from which new stars are born. By measuring elemental abundance ratios in stellar atmospheres, we can infer the composition of the cloud of dust and gas from which the stars were born and the nucleosynthetic pathways that could have produced such compositions \citep{Johnson2019}. This data, compared to Galactic Chemical Evolution (GCE) models based on empirical calibrations, enables us to infer stellar ages.

[Y/Mg], the yttrium (Y) to magnesium (Mg) abundance ratio of a star relative to the Sun, has been proposed as an age diagnostic for solar-type stars \citep{daSilva2012,Nissen2015,Maia2016}. As an $\alpha$-element, Mg is produced primarily by the core-collapse supernovae of massive stars ($\mathrm{M}_\star$ $>$ 8 $\mathrm{M}_\odot$) with minor contributions from intermediate mass stars \citep[2 $<$ $\mathrm{M}_\star$ $<$ 8 $\mathrm{M}_\odot$,][]{Vangioni2019}. Massive stars burn through their hydrogen much more quickly than their low-mass counterparts, and hence are expected to have populated the interstellar medium with larger abundances of $\alpha$-elements at earlier times. In contrast, intermediate-mass stars produce elements such as Y through the slow neutron capture process (s-process) during their asymptotic giant branch phase. This is shortly before they expel their envelopes into the surrounding interstellar medium. Because intermediate-mass stars have longer evolutionary timescales due to their lower hydrogen burning rates, enrichment of s-process elements in the interstellar medium and eventual stellar atmospheres is expected to occur at later times in our Galaxy's evolution relative to the $\alpha$-elements. Therefore, the principles of GCE predict that [Y/Mg] has increased with time such that it decreases with stellar age. 

According to \citet{Nissen2015}, \citet{Maia2016}\labeltext{TM16}{TM16} (hereafter \ref{TM16}), and \citet{Spina2016}, [Y/Mg] is a precise clock for solar-twins, defined as stars within $\pm$100 K of solar \teff, and within $\pm$0.1 dex in \logg\ and metallicity [Fe/H] \citep{Ramirez2014}. \ref{TM16} shows that differential measurements of [Y/Mg] can produce ages as precise as 0.8 Gyr for solar twins. More recent work has corroborated the [Y/Mg] ``Clock'' for solar-type stars \citep{Anders2018} and those with \teff\ between 5700 and 6400 K \citep{Nissen2017}, evolved solar metallicity stars \replaced{\citep{Slumstrup2017}}{\citep{Slumstrup2017,Casamiquela2021}}, thin and thick disk stars \citep{Titarenko2019}, and stars in the solar- and outer-disk regions of the Galaxy \citep{Vazquez2022}. However, \citet{Feltzing2017} find that the [Y/Mg] Clock may only be useful within a narrow range of [Fe/H], and show that the relation is effectively flat for metallicities of $\approx$ --0.5 dex, and \citet{Vazquez2022} show that the Clock changes as a function of galactocentric distance. Similarly, while \citet{Titarenko2019} also find a tight correlation between [Y/Mg] and age, a difference in slopes for thin and thick disk stars suggests the [Y/Mg] Clock is not universal and instead Galactic-neighborhood-dependent. Therefore, it is unclear how useful the [Y/Mg] Clock is for the wider range of \teff, \logg, and [Fe/H] present in field stars. Here, we use the well-studied \kep\ \citep{borucki10} planet host stars with high-resolution spectra and rotation periods to test this relation.

\section{Stellar Sample}
\label{sec:sample}

\begin{figure}
    \centering
    \resizebox{\hsize}{!}{\includegraphics{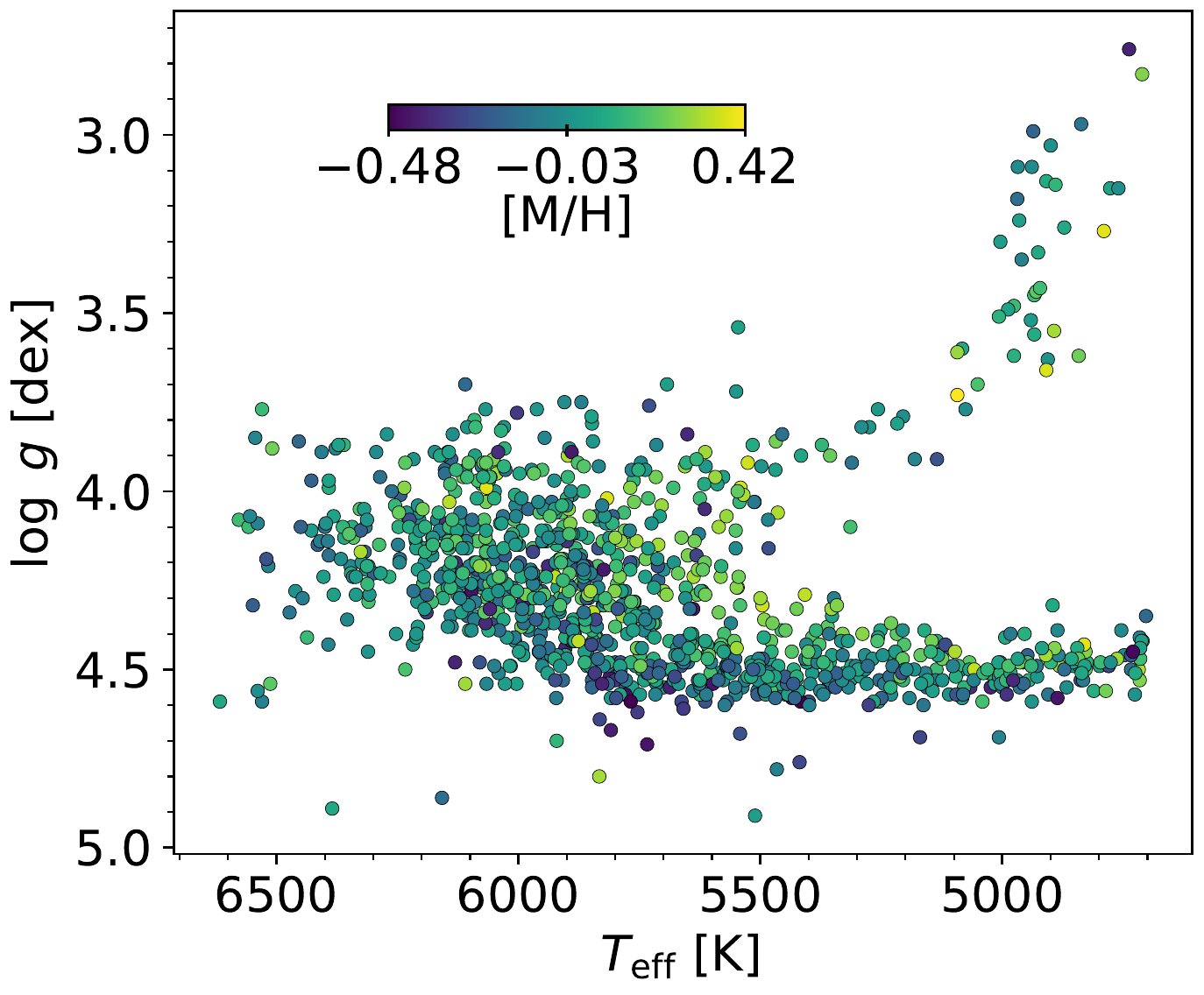}}
    \caption{Kiel diagram of 1100 $Kepler$ planet host stars with self-consistent spectroscopic \teff, \logg, [M/H], [Y/Mg], and age measurements from \citet{Brewer2018}. Stars are colored according to spectroscopic [M/H] as determined by \citet{Brewer2018}.}
    \label{fig:ymgHR}
\end{figure}

In Figure \ref{fig:ymgHR}, we plot the \kep\ host star sample with measured \teff, \logg, metallicity, yttrium abundance (A(Y)), and magnesium abundance (A(Mg)) from \citet{Brewer2018}. \citet{Brewer2018} used Spectroscopy Made Easy \citep[SME,][]{SME} to derive the stellar atmospheric parameters and abundances. We removed three stars with metallicities below --0.5 dex (poor statistics). The remaining 1100 FGK stars have metallicities ranging from --0.48 to 0.42 dex and most are dwarfs, although a few have started evolving up the giant branch. Unlike the solar twin samples from \citet{Nissen2015}, \ref{TM16}, and \citet{Spina2016}, this plot includes both more evolved stars and stars of spectral types F and K.

\begin{figure*}[t!]
    \centering
    \resizebox{\hsize}{!}{\includegraphics{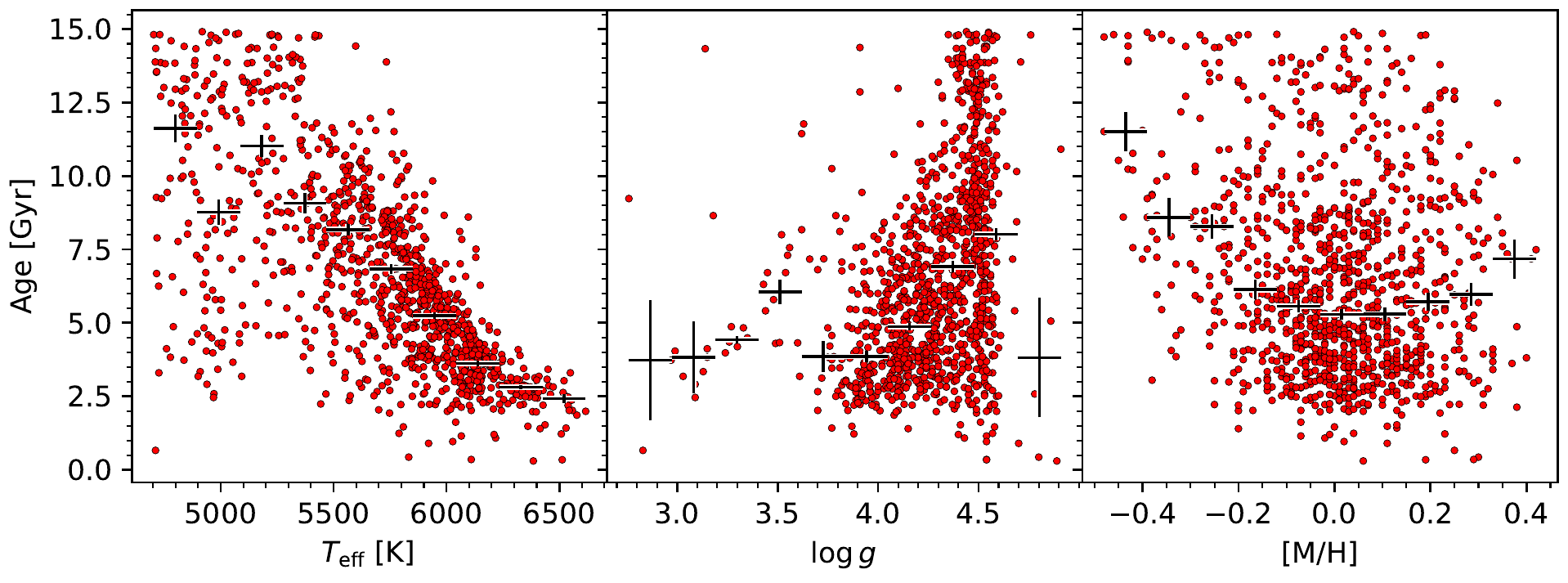}}
    \resizebox{\hsize}{!}{\includegraphics{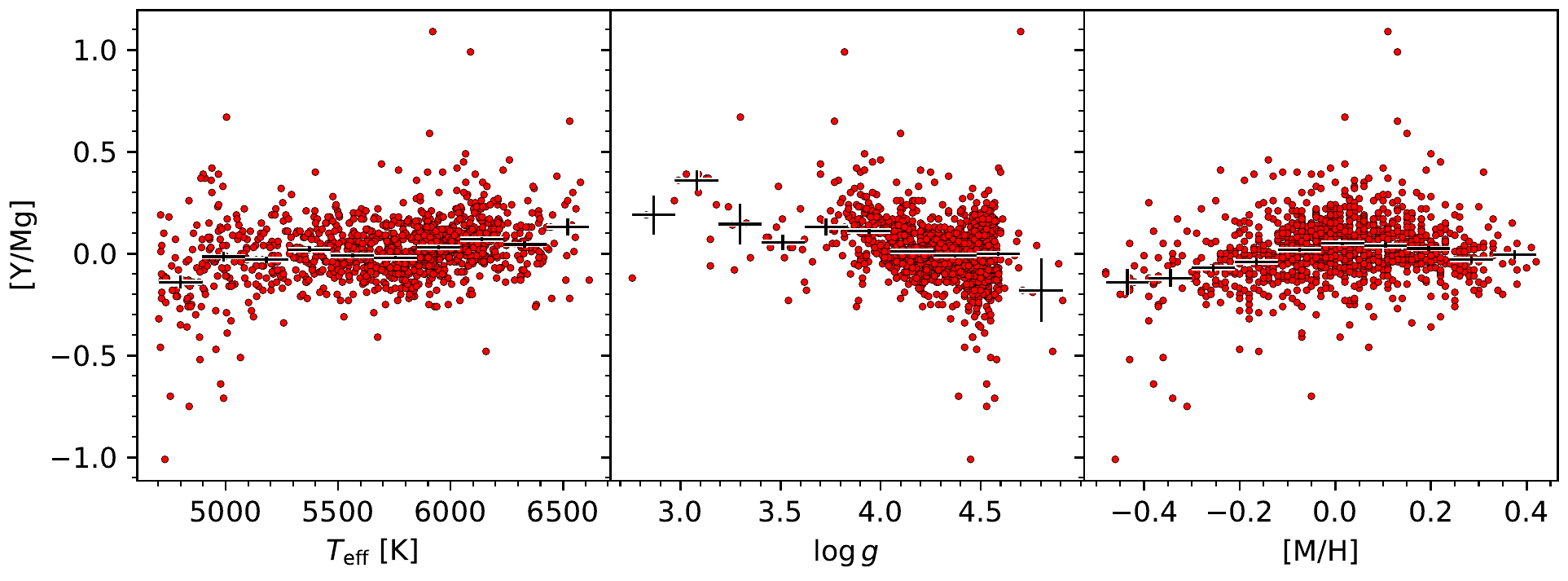}}
    \caption{$Top$: Age as a function of \teff, \logg, and metallicity for the \ref{BF18} sample. Median ages for each bin are shown as crosses, and vertical bars represent the standard error of the mean for each bin. $Bottom$: [Y/Mg] as a function of the spectroscopic parameters for the \ref{BF18} sample. We omitted stars with metallicities $<$ --0.5 because of their sparseness.}
    \label{fig:specParComp}
\end{figure*}

In addition to [Y/Mg] measurements, we also adopt the stellar ages from \citet{Brewer2018}\labeltext{BF18}{BF18} (hereafter \ref{BF18}), which are self-consistent with the derived spectroscopic parameters and elemental abundances. \ref{BF18} used \texttt{isochrones} \citep{morton15} to infer fundamental stellar parameters including age from a combination of the spectroscopic parameters detailed above, \gaia\ DR2 parallaxes \citep{Brown2018}, 2MASS $K_s$ magnitudes \citep{skrutskie06}, and Dartmouth Stellar Evolution Database models \citep{dotter08}. \citet{Fulton2018} also inferred fundamental parameters for the same stellar sample, but they did not measure [Y/Mg], the central abundance ratio of this work. Therefore, we choose to use the inherently self-consistent atmospheric parameters, abundances, and stellar ages of \ref{BF18} to minimize potential systematics between data sets. We will also compare the asteroseismic ages determined for the 34 $Kepler$ planet host stars in common with \ref{BF18} from \citet{Aguirre2015} and \citet{Creevey2017}.

\section{[Y/Mg] and Age as a Function of Spectroscopic Parameters}

An important first step is to test whether [Y/Mg] correlates with spectroscopic parameters independently of age. Figure \ref{fig:specParComp}, shows age as a function of \teff, \logg, and metallicity. As expected, ages increase for cool stars with lower surface gravities and more metal-poor stars. Additionally, the age scatter is largest for cool main sequence stars. This behavior is expected due to the strong relationship between stellar mass and stellar lifetimes.

The bottom panels of Figure \ref{fig:specParComp} show [Y/Mg] versus the spectroscopic parameters. Compared to the trends seen in the age plots, the [Y/Mg] trends shown here are weaker yet still significant. In particular, we see the smallest [Y/Mg] abundances at low \teff\ and the largest [Y/Mg] abundances at high \teff, although around solar \teff\ the median bins do not vary much. Most of the differences in median binned \teff\ occur for the coolest and hottest stars, although they do have larger uncertainties given their smaller relative number. We see [Y/Mg] increase for \logg\ below 4.0 dex, and peak at [M/H] $\approx$ 0.0 dex, with gradual and/or insignificant trends otherwise. In addition, we note that some natural systematic correlations are expected if [Y/Mg] is related to age, given how age is correlated with \teff\ and \logg\ through the stellar lifetime, and how overall metallicity should differ as a function of [Y/Mg], given their similar observables. Ultimately, the systematics shown here are important to consider when interpreting trends of [Y/Mg] a function of stellar age.

\section{[Y/Mg] as a Function of Stellar Age}
\label{sec:ymgage}

\begin{figure*}
    \centering
    \resizebox{0.85\hsize}{!}{\includegraphics{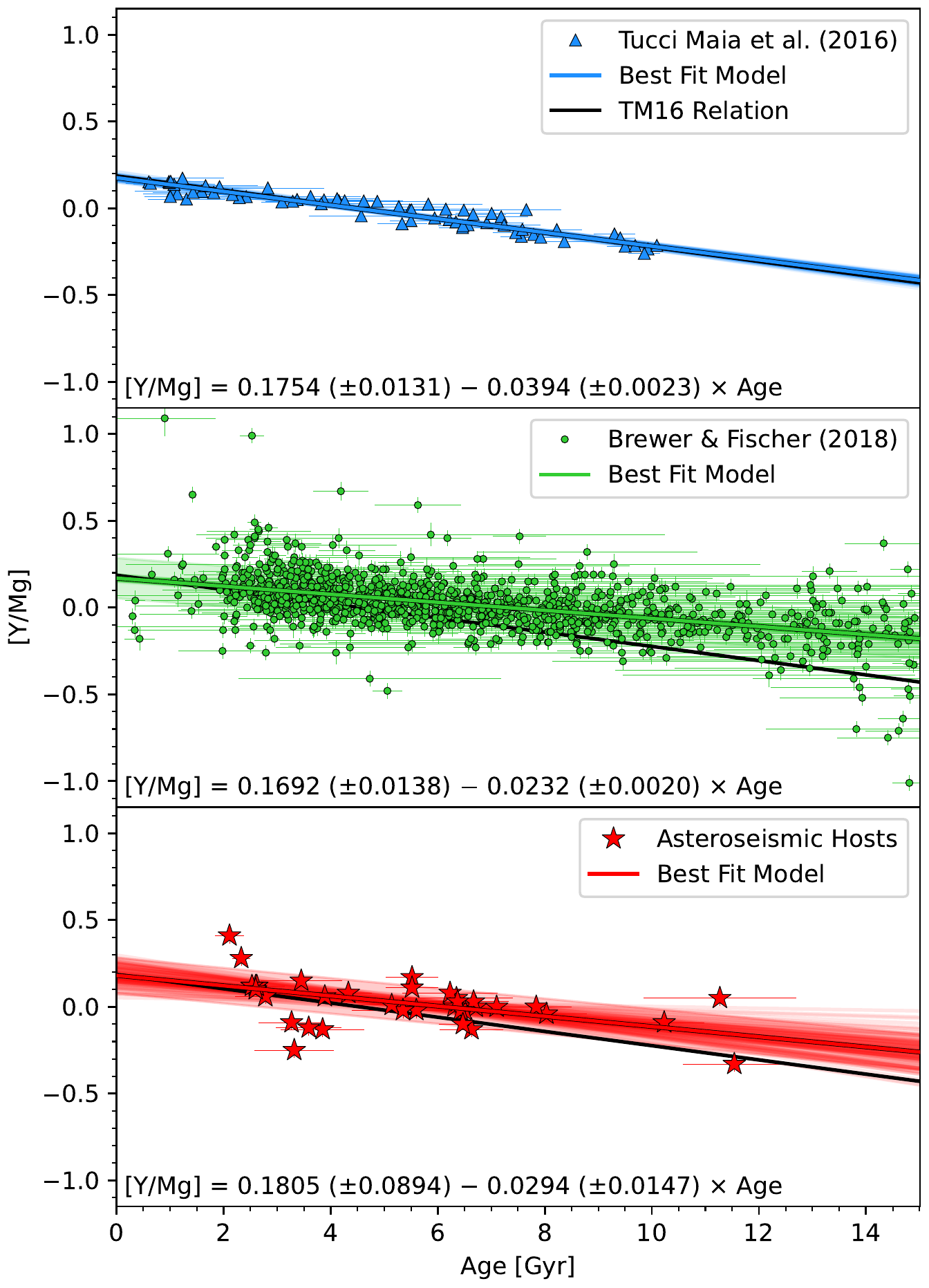}}
    \caption{[Y/Mg] versus stellar age for three samples:  (1) \ref{TM16} solar twins with solar-differential [Y/Mg] abundances from the \citet{Ramirez2014} analysis of MIKE spectra \citep{Bernstein2003} and Yonsei-Yale isochrone ages \citep{yi01} ($top$), (2) \ref{BF18} $Kepler$ planet host stars with [Y/Mg] abundances and isochrone ages based on Dartmouth Stellar Evolution Database \citep[DSEP,][]{dotter08} ($middle$), and (3) $Kepler$ planet host stars with asteroseismic ages from \citet{Aguirre2015} or \citet{Creevey2017} and \cite{Brewer2018} [Y/Mg] abundances ($bottom$). Individual MCMC realizations are shown as the color-matched translucent curves and the intrinsic scatters are shown as the lightly shaded regions surrounding the best fit model (where visible). In black is the best-fit expression from \ref{TM16}, and uncertainties are plotted for each individual star.}
    \label{fig:ymgagemcmc}
\end{figure*}

Figure \ref{fig:ymgagemcmc} compares [Y/Mg] versus age for the \ref{TM16}, \ref{BF18}, and asteroseismic host star samples. While all data sets show [Y/Mg] decreasing with age, the \kep\ host star \ref{BF18} and asteroseismic correlation coefficients are weaker -- $R^2$ = 0.27 and 0.23, respectively, versus 0.89 for \ref{TM16} -- with larger scatter -- $\sigma$ = 0.14 and 0.12 dex, respectively versus 0.04 dex for \ref{TM16}. At least some of the additional scatter can be explained by the difference in typical [Y/Mg] uncertainties (0.02 dex for \ref{TM16} versus 0.05 dex for \ref{BF18}). This difference occurs because \ref{TM16} performed differential abundance analyses with respect to the sun, an approach only valid for solar twins, whereas \ref{BF18} did not. Still, the 0.1 dex difference in scatter is not covered by the 0.03 dex difference in typical [Y/Mg] uncertainties, which means there either must be astrophysical scatter introduced by the diverse FGK hosts in \ref{BF18} or additional systematics. The formal uncertainties on the isochrone ages reported by both \ref{BF18} and \ref{TM16} appear to be similar to those of the asteroseismic ages, which is likely due to the small \teff\ uncertainties used in both isochrone analyses. Systematics in the reported ages are likely much larger \citep{Tayar2022}. In addition, the \ref{BF18} hosts display a larger scatter in [Y/Mg] at the oldest ages.

\ref{TM16} report the following best-fit linear relation,
\begin{equation}\label{eq:tm}
    [\mathrm{Y/Mg}] = 0.184(\pm 0.008) - 0.041 (\pm 0.001) \times \mathrm{Age},
\end{equation}where the Age is in Gyr. Given the uncertainties on the best-fit slope and intercept, this implies a tight correlation between age and [Y/Mg]. To test the robustness of this result, we performed linear fits while accounting for x- and y- uncertainties using \texttt{emcee} \citep{Foreman-Mackey2013} and implemented the following likelihood function \citep{Hogg2010}:
\begin{equation}\label{eq:likelihood}
    \ln{\mathcal{L}} = -\frac{1}{2} \sum_{i=1}^N \ln{\left(\Sigma_i^2 + V\right)} -\frac{1}{2} \sum_{i=1}^N \frac{\Delta_i^2}{\Sigma_i^2 + V},
\end{equation}where
\begin{equation}\label{eq:deltasigmav}
\begin{split}
    \Delta_i & = \hat{\vect{v}}^\top \vect{Z}_i - b \cos{\theta}, \\
    \Sigma_i & = \hat{\vect{v}}^\top \vect{S}_i \hat{\vect{v}}, \\
    V & = c^2 \cos^2{\theta},
\end{split}
\end{equation}and where
\begin{equation}\label{eq:vectors}
\begin{split}
    \hat{\vect{v}} = \begin{bmatrix}
                     - \sin{\theta} \\
                     \cos{\theta} \end{bmatrix},\
    \vect{Z}_i = \begin{bmatrix}
                     x_i \\
                     y_i \end{bmatrix},\
    \vect{S}_i = \begin{bmatrix}
                    \sigma_{xi}^2 & \sigma_{xi}\sigma_{yi} \\
                    \sigma_{xi}\sigma_{yi} & \sigma_{yi}^2 \end{bmatrix},
\end{split}
\end{equation}with the slope $m$ parameterized as $\theta = \arctan{(m)}$ and best fit model parameterized as $b \cos{\theta}$ to treat all possible slopes equally. $\Delta$ represents the residuals and $\Sigma$ is the covariance rotated parallel to the slope. $c$ is a measure of the point-to-point scatter not contained within the formal uncertainties in units of dex, and we may refer to it as the intrinsic scatter from now on.  

We used 32 walkers with 20000-step chains in the three-dimensional parameter space with uniform priors on the slope (--1 $<$ $m$ $<$ 1 dex/Gyr), intercept (--2 $<$ $b$ $<$ 2 dex), and additional [Y/Mg] scatter (--20 $<$ $\ln{c}$ $<$ 5) as our free parameters. We used \texttt{emcee}'s \texttt{get}\_\texttt{autocorr}\_\texttt{time}() function to ensure the fits converged within the first 400 steps of sampling. We then identified the longest autocorrelation time of our three free parameters and removed four times that number of steps to account for burn-in. Examples of the best fit relations for each data set are shown in Figure \ref{fig:ymgagemcmc}. We also ran 1000 bootstrap simulations to quantify the impact of outliers on the fit parameters. In particular, we ran each bootstrap simulation by drawing, with replacement, the same number of stars as the observed sample and then used \texttt{scipy}'s \texttt{minimize} function to determine the maximum likelihood fit parameters from the likelihood in Equation \ref{eq:likelihood}. We then computed the standard deviation of the 1000 bootstrapped slopes and intercepts and then added them in quadrature to the MCMC uncertainties to produce our reported uncertainties:
\begin{equation}\label{eq:sigmas}
\begin{split}
    \sigma_m = \sqrt{\sigma_{m,\mathrm{MCMC}}^2 + \sigma_{m,\mathrm{BOOT}}^2},\\
    \sigma_b = \sqrt{\sigma_{b,\mathrm{MCMC}}^2 + \sigma_{b,\mathrm{BOOT}}^2}.
\end{split}
\end{equation}
where $\sigma_{\mathrm{MCMC}}^2$ and $\sigma_{\mathrm{BOOT}}^2$ represent uncertainties derived from the MCMC and bootstrap analyses, respectively.

The top panel in Figure \ref{fig:ymgagemcmc} shows our best-fit relation for the \ref{TM16} sample. We compute a slope of $m$ = --0.0394 $\pm$ 0.0023 dex/Gyr and an intercept of $b$ = 0.175 $\pm$ 0.013 dex, which matches Equation \ref{eq:tm} within uncertainties. This result is in statistical agreement with the reported relation of \ref{TM16} and the slope is still significant at $\approx$17$\sigma$, despite the differences in our methods.

The middle panel in Figure \ref{fig:ymgagemcmc} displays the $Kepler$ planet host star sample with stellar ages from \ref{BF18}. Our best fit relation for the 1100-star sample has a slope $m$ = --0.0232 $\pm$ 0.0020 dex/Gyr, significant at $\approx$12$\sigma$, which is shallower and has significantly more intrinsic scatter ($c$ = 0.12 dex) than the \ref{TM16} relation. The shallow slope computed here suggests that the [Y/Mg] Clock may not be as strong an age diagnostic as previously reported, at least for the wide range of \teff, \logg, and metallicities present in the \ref{BF18} \kep\ host star sample. In addition, the large $c$ here indicates there is true astrophysical and/or systematic scatter introduced by the diverse FGK star sample. 

The bottom panel of Figure \ref{fig:ymgagemcmc} shows the best fit relation for the asteroseismic stars from \citet{Aguirre2015} and \citet{Creevey2017}. The slope, at $m$ = --0.029 $\pm$ 0.015 dex/Gyr, is consistent with the \ref{BF18} and \ref{TM16} relations, as the smaller sample size and large intrinsic scatter ($c$ = 0.10 dex) produces a more uncertain [Y/Mg] Clock when compared to the significantly larger \ref{BF18} and tighter \ref{TM16} samples. We compile our best-fit [Y/Mg]-Age relations and uncertainties in Table \ref{tab:agefits}.

To determine each relation's subsequent uncertainty in stellar age, we performed similar MCMC analyses where we define age as the ordinate and [Y/Mg] as the abscissa, with uniform priors on the slope (--100 $<$ $m$ $<$ 10 Gyr/dex), intercept (--10 $<$ $b$ $<$ 100 Gyr), and intrinsic scatter (--20 $<$ $\ln{c}$ $<$ 3). The subsequent best-fit relations produce residuals with scatters of 0.95, 5.8, and 6.1 Gyr in age for the \ref{TM16}, \ref{BF18}, and asteroseismic samples, respectively.

\begin{figure*}
    \centering
    \resizebox{\hsize}{!}{\includegraphics{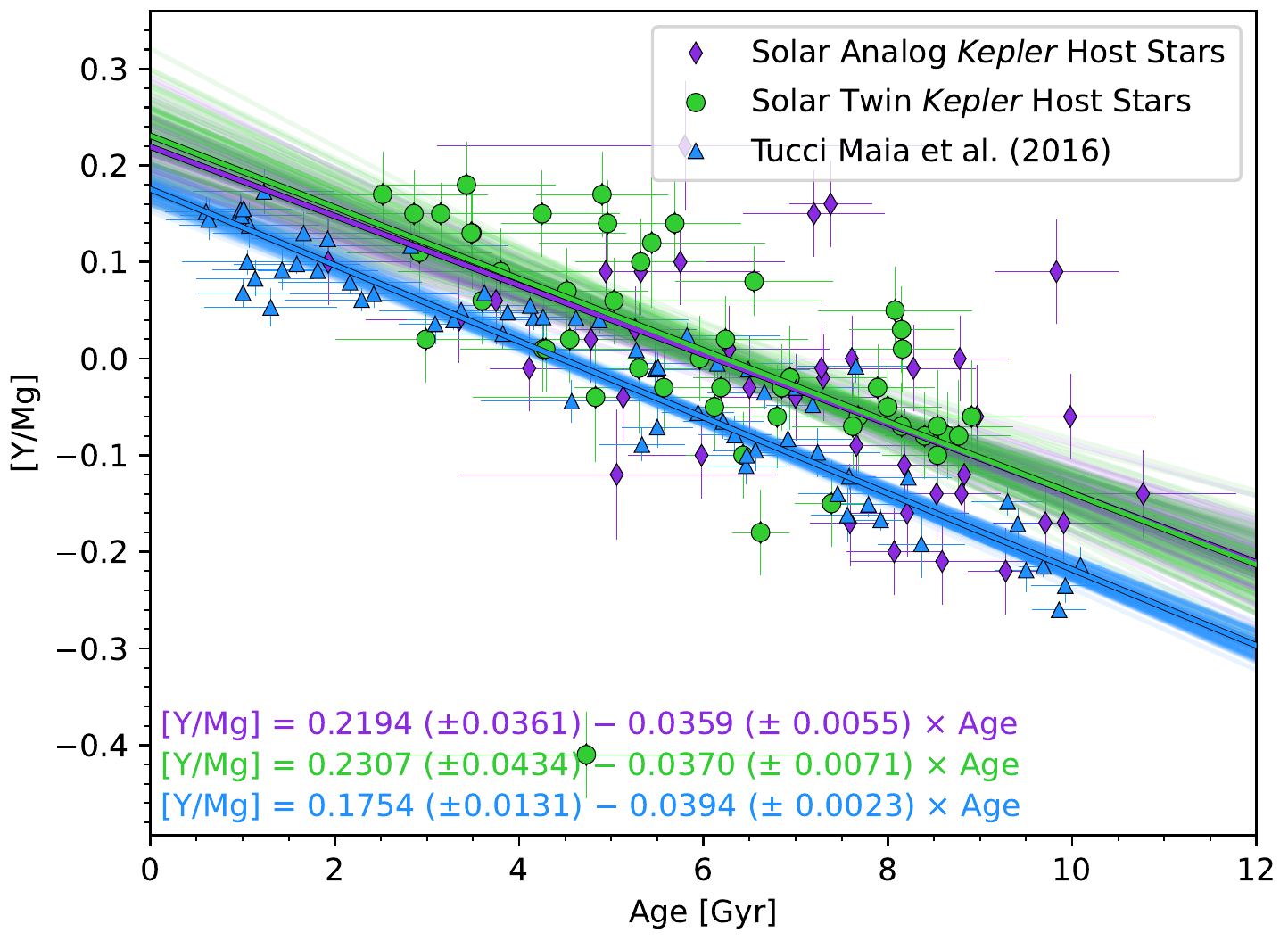}}
    \caption{[Y/Mg] versus age for $Kepler$ solar twins (green circles), defined as stars within $\pm$100 K in \teff\ and $\pm$0.1 dex in \logg\ and [M/H] relative to the Sun. Solar analogs (purple diamonds) have similar \teff\ and \logg\ as the Sun, but [M/H] is allowed to vary. Color-matched translucent lines show random MCMC samples. Solar twins are a subset of the solar analogs by definition, and hence are plotted on top. The \ref{TM16} relation and corresponding data are plotted in blue.}
    \label{fig:ymgtwin}
\end{figure*}

We also performed $F$-tests using \texttt{scipy}'s \texttt{stats.f} to determine the significance at which a two parameter (slope + intercept) model is better than a one parameter (intercept-only) model. The intercept-only model assumes a flat slope and hence no [Y/Mg]-Age dependence. For the \ref{TM16}, \ref{BF18}, and asteroseismic data, we compute p-values consistent with 12.4$\sigma$, 18.6$\sigma$, and 2.9$\sigma$ significances, respectively. Therefore, the \ref{TM16} and \ref{BF18} data statistically prefer the slope + intercept model at $>>$3$\sigma$ significance over the intercept-only model, while the asteroseismic data prefer the slope + intercept model with marginal significance due to their small sample size and large scatter. While we acknowledge the large discrepancy between the 12$\sigma$ $F$-test and the 17$\sigma$ slope for the \ref{TM16} data, we note that the magnitude of [Y/Mg] uncertainties significantly affects the precision of the MCMC-derived parameters while it has little-to-no impact on the $F$-test results.

\subsection{The [Y/Mg] Clock for $Kepler$ Solar-Twins and Analogs}
\label{sec:ymgtwins}

We now isolate a subsample of \ref{BF18} stars that is consistent with solar twins and analogs. We use the same definition of solar twin as \ref{TM16} and \citet{Ramirez2014}: $\pm$100 K in \teff\ and $\pm$0.1 dex in \logg\ and [M/H]. Because our sample's uncertainties are larger than those of \ref{TM16}, our solar-twins and analogs will be more contaminated by non-twin stars, which may produce flatter [Y/Mg]-Age relations. We found no solar twins with asteroseismic constraints in our overlapping sample.

Figure \ref{fig:ymgtwin} shows that the \ref{BF18} solar twins, compared to \ref{TM16}, produce a statistically similar slope (--0.0370 $\pm$ 0.0071 dex/Gyr versus --0.0394 $\pm$ 0.0023 dex/Gyr), a $\gtrsim$1$\sigma$ larger intercept (0.231 $\pm$ 0.043 dex versus 0.175 $\pm$ 0.013 dex), and similar intrinsic scatters (both less than 0.001 dex). Unsurprisingly, the \ref{BF18} solar twin [Y/Mg]-Age relation is more similar to \ref{TM16} compared to the full \ref{BF18} sample in Figure \ref{fig:ymgagemcmc}. When using different [Y/Mg]-Age relations, a significant intercept offset for relations with equivalent slopes will also produce significantly different ages. We compute a 2.6 Gyr uncertainty in age from the \ref{BF18} solar twin relation, which is larger than the 0.95 Gyr uncertainty in the \ref{TM16} relation. We note that differences in slopes, intercepts, and age uncertainties may arise from any combination of factors: (1) the \ref{BF18} data are both more uncertain and may contain non-solar twin contaminants, (2) \ref{TM16} used a differential abundance analysis while \ref{BF18} did not, and/or (3) \ref{TM16} and \ref{BF18} used different models to determine stellar age, which can easily produce age offsets \citep{Tayar2022}. Because we allow for intrinsic scatter, the outlier at age $\approx$ 5 Gyr and [Y/Mg] $\approx$ --0.4 dex does not affect our best fit line significantly.

The purple line in Figure \ref{fig:ymgtwin} shows our best fit relation to the \ref{BF18} solar analogs, defined as stars within $\pm$100 K in \teff\ and $\pm$0.1 dex in \logg\ relative to the Sun but with varying metallicities. We compute a slope that is statistically indistinguishable from the \ref{TM16} relation as we did for solar twins, but the solar analogs exhibit a larger intrinsic scatter (0.001 dex) than both the \ref{BF18} solar twins and \ref{TM16} data. The [Y/Mg]-Age scatter corresponds to an uncertainty in age of 2.6 Gyr, similar to the \ref{BF18} solar twin age uncertainty. Our measured intercept is $\gtrsim$1$\sigma$ larger than the \ref{TM16} intercept as was found for the solar twin sample above. We also note that choosing solar analogs with super- or sub-solar metallicities results in a shallower [Y/Mg]-Age slope.

To estimate typical systematic age uncertainties given [Y/Mg] for solar twins, we added our best-fit [Y/Mg]-Age results to those of the literature \citep{Nissen2015,Nissen2016,Spina2016,Maia2016,Nissen2017} compiled in Table 6 of \citet{Delgado2019}. We computed a mean slope of --0.0389 $\pm$ 0.0025 dex/Gyr and a mean intercept of 0.185 $\pm$ 0.025 dex, where the uncertainties are based on the standard deviation of the various estimates. From these estimates, we then propagated our uncertainties on the slope, intercept, and [Y/Mg] measurements (0.05 dex from \ref{BF18}) to produce a 1.5 Gyr systematic uncertainty.

We conducted $F$-tests to compare one- and two-parameter models as before. For the \ref{TM16} solar twins, \ref{BF18} solar twins, and the \ref{BF18} solar analogs, we compute p-values consistent with 12.4$\sigma$, 4.2$\sigma$, and 6.1$\sigma$, respectively. Therefore, all three data sets statistically prefer the slope + intercept model at $>$4$\sigma$ significance.

\subsection{A More Reliable [Y/Mg] Clock Sample}
\label{sec:reliablesamp}

\begin{figure}
    \centering
    \resizebox{\hsize}{!}{\includegraphics{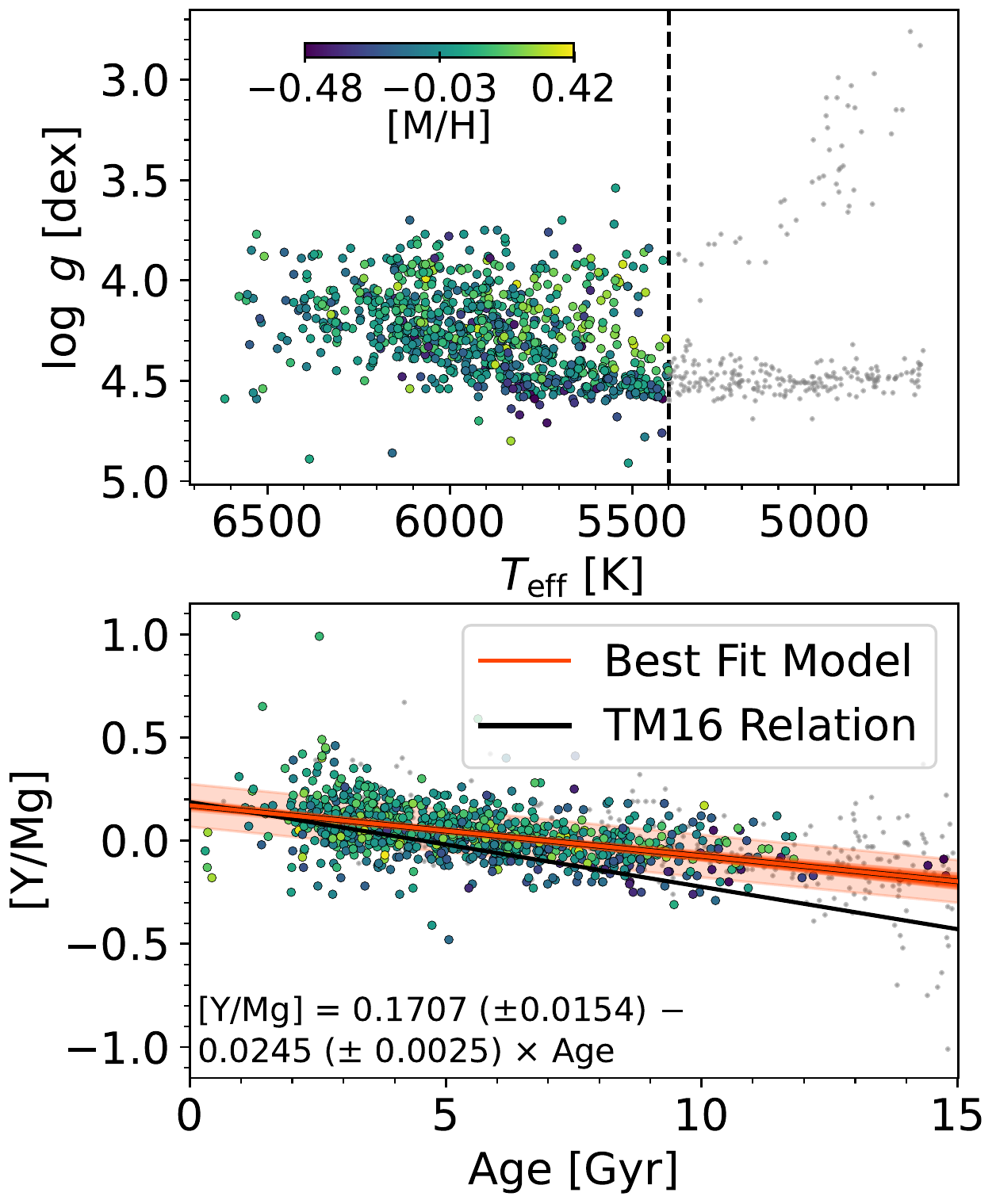}}
    \caption{$Top$: Kiel Diagram of 1100 $Kepler$ planet host stars with self-consistent spectroscopic \teff, \logg, [M/H], [Y/Mg], and age measurements from \ref{BF18}. Eight hundred and forty-six stars with informative ages (\teff\ $>$ 5400 K) are colored according to their spectroscopic [M/H] determined by \ref{BF18}, while all other stars are represented by small grey circles. $Bottom$: [Y/Mg] versus stellar age, where stars are colored equivalently as in the top panel; the MCMC best fit is plotted in orange-red for the 846 stars with reliable ages.}
    \label{fig:ymgHRAge}
\end{figure}

Some of the increased scatter of the \ref{BF18} sample could be due to uncertainties in stellar ages. We therefore removed both lower-main sequence stars and giant stars by cutting all stars with \teff\ $\leq$ 5400 K. The lower-main sequence stars have the most uncertain isochrone ages due to their slow evolution on the main sequence \citep{Berger2020a,Berger2020b}, while the systematics in giant star ages are likely larger due to their strong dependence on model input physics \citep{Tayar2022}.

The 846-star sample is shown in Figure \ref{fig:ymgHRAge}. Our MCMC analysis produces $m$ = --0.0245 $\pm$ 0.0025 dex/Gyr, $b$ = 0.171 $\pm$ 0.015 dex, and an intrinsic scatter $c$ = 0.10 dex. The slope is statistically consistent with the full sample's slope of $m$ = --0.0232 $\pm$ 0.0020 dex/Gyr and $\approx$4$\sigma$ more shallow than the \ref{TM16} slope. In addition, we determined an uncertainty of 5.0 Gyr in age compared to the full sample's uncertainty of 5.8 Gyr. We computed an $F$-test p-value significance of 14.5$\sigma$, which indicates that a slope + intercept model is strongly preferred over the intercept-only model. We do not see large slope/intercept/$c$ differences in the reduced sample compared to the full sample above because we retain the majority of stars from the full sample, and the stars that are removed, while concentrated at old stellar ages with large variations in [Y/Mg], occur roughly as frequently above and below the previous best-fit trendline. We suspect these large [Y/Mg] variations at old age arise either from difficulty in measuring [Y/Mg] and/or because the majority of the removed stars are low-mass dwarfs with uninformative isochrone ages due to their slow evolution in the HR diagram.

\subsection{The [Y/Mg] Clock and Stellar \teff}
\label{sec:ymgteff}

\begin{figure}
    \centering
    \resizebox{\hsize}{!}{\includegraphics{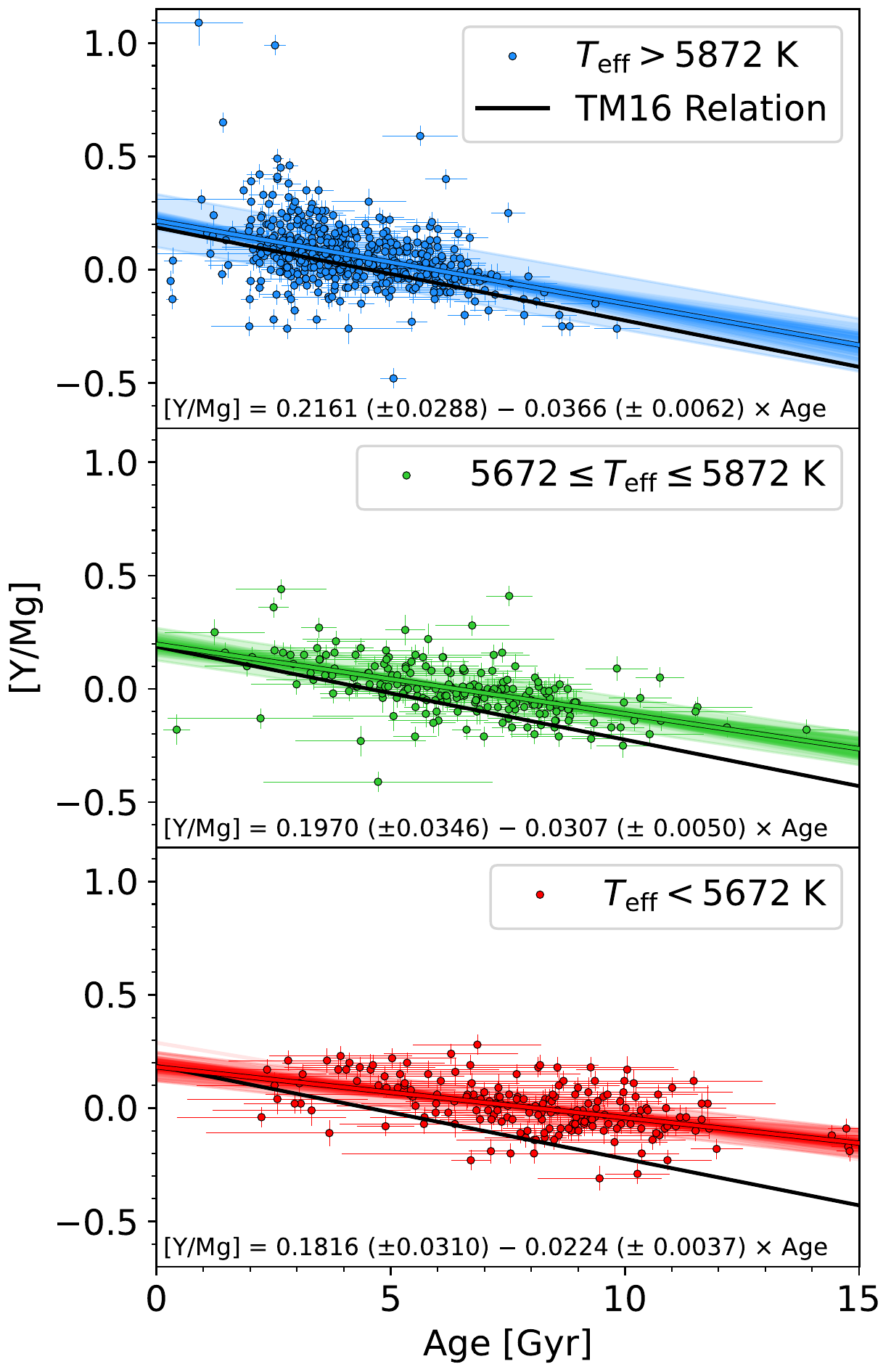}}
    \caption{[Y/Mg] versus stellar age for the reduced \ref{BF18} sample. The top, middle, and bottom panels show \teff\ $>$ 5872 K, \teff\ between 5672 and 5872 K, and \teff\ $<$ 5672 K, respectively. We plot the best fitting relations as solid lines and the MCMC realizations as translucent lines underneath. In black we plot the \ref{TM16} best fit relation.}
    \label{fig:ymgAgeTeff}
\end{figure}

Figure \ref{fig:ymgAgeTeff} shows the [Y/Mg] Clock for the \ref{BF18} sample split into three \teff\ bins: (1) hotter, (2) solar, and (3) cooler stars. We observe that solar \teff\ stars exhibit the least scatter and the most statistically significant slope with a corresponding age uncertainty of 3.5 Gyr, while the cooler stars have similar scatter with the shallowest slope and a corresponding age uncertainty of 4.5 Gyr. Comparatively, hotter stars exhibit the most scatter and produce the steepest slope with a corresponding age uncertainty of 3.8 Gyr. $F$-test results suggest that all three samples strongly prefer the slope + intercept model over the intercept-only fit. \replaced{Exact fit parameters can be found in Table \ref{tab:agefits}}{Table \ref{tab:agefits} lists the best fit parameters and uncertainties}.

We do not observe any strong trends in the [Y/Mg] Clock as a function of \teff, and find slopes and intercepts in the hotter, solar, and cooler \teff\ samples that are statistically consistent with one another, except for the $>$1$\sigma$ shallower slope for cooler \teff\ stars relative to solar \teff\ stars.

\subsection{The [Y/Mg] Clock and Stellar Surface Gravity}
\label{sec:ymglogg}

\begin{figure}
    \centering
    \resizebox{\hsize}{!}{\includegraphics{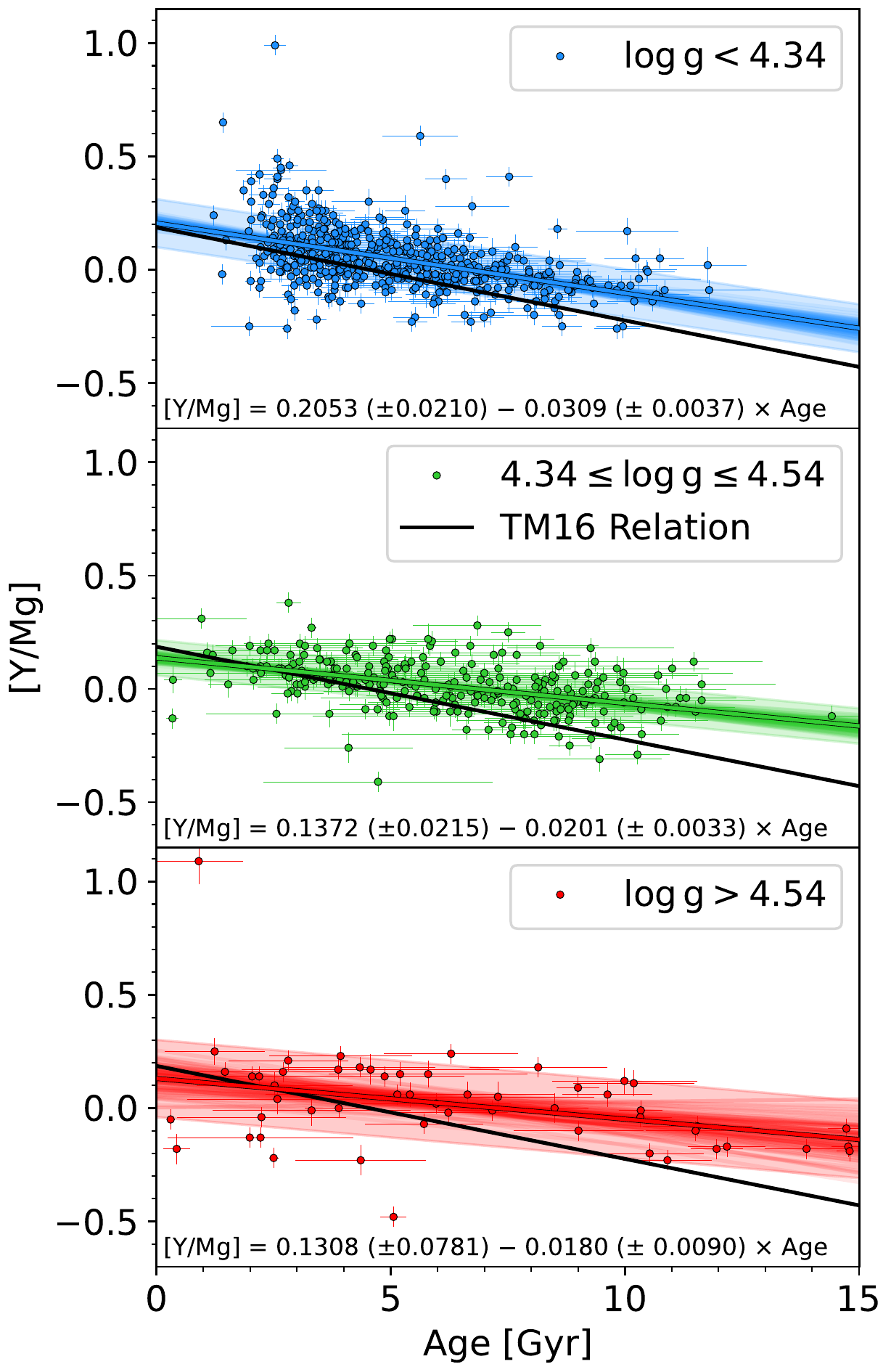}}
    \caption{[Y/Mg] versus stellar age for the reduced \ref{BF18} sample. The top, middle, and bottom panels show \logg\ $<$ 4.34 dex, \logg\ between 4.34 and 4.54 dex, and \logg\ $>$ 4.54 dex, respectively. We plot the best fitting relations as solid lines and the MCMC realizations as translucent lines underneath. In black we plot the \ref{TM16} best fit relation.}
    \label{fig:ymgAgelogg}
\end{figure}

Similar to \teff, \logg\ provides another dimension in which to test the scope of the [Y/Mg] Clock. For instance, \citet{Slumstrup2017} used open cluster data to show that the [Y/Mg]-Age relation derived for solar twins in \cite{Nissen2015} holds for solar-metallicity giant stars in the helium-core-burning phase. Therefore, we test the relation for the \kep\ dwarf and subgiant stars studied here.

Figure \ref{fig:ymgAgelogg} shows the [Y/Mg] Clock for stars of lower, solar, and higher \logg. The lower \logg\ stars produce the statistically steepest [Y/Mg]-Age slope with an intrinsic scatter of 0.11 dex, while the solar and higher \logg\ stars produce statistically consistent slopes that are $>$1$\sigma$ more shallow than the lower \logg\ stars and exhibit intrinsic scatters of 0.078 and 0.17 dex, respectively. For the lower, solar, and higher \logg\ stars, the corresponding age uncertainties are 3.9, 5.3, and 14 Gyr, and $F$-test p-value significances are 11.8, 7.9, and 3.0$\sigma$, respectively. The 14 Gyr uncertainty on the higher \logg\ stars is largely due to a combination of the large intrinsic scatter and the 2$\sigma$ shallow slope. Summaries of the fit parameters can be found in Table \ref{tab:agefits}.

Ultimately, our \logg\ results suggest that the most sensitive [Y/Mg]-Age relationship occurs for the lower \logg\ stars, which are the largest of the three \logg\ samples. In general, isochrones are most sensitive for these stars and may result in a more precisely measured relationship. We also measure the most uncertain relation for higher \logg\ stars, which comprise the smallest \logg\ sample, produce the largest age and intrinsic scatter, and the smallest $F$-test significance.

\subsection{The [Y/Mg] Clock and Stellar Metallicity}
\label{sec:ymgmet}

\begin{figure}
    \centering
    \resizebox{\hsize}{!}{\includegraphics{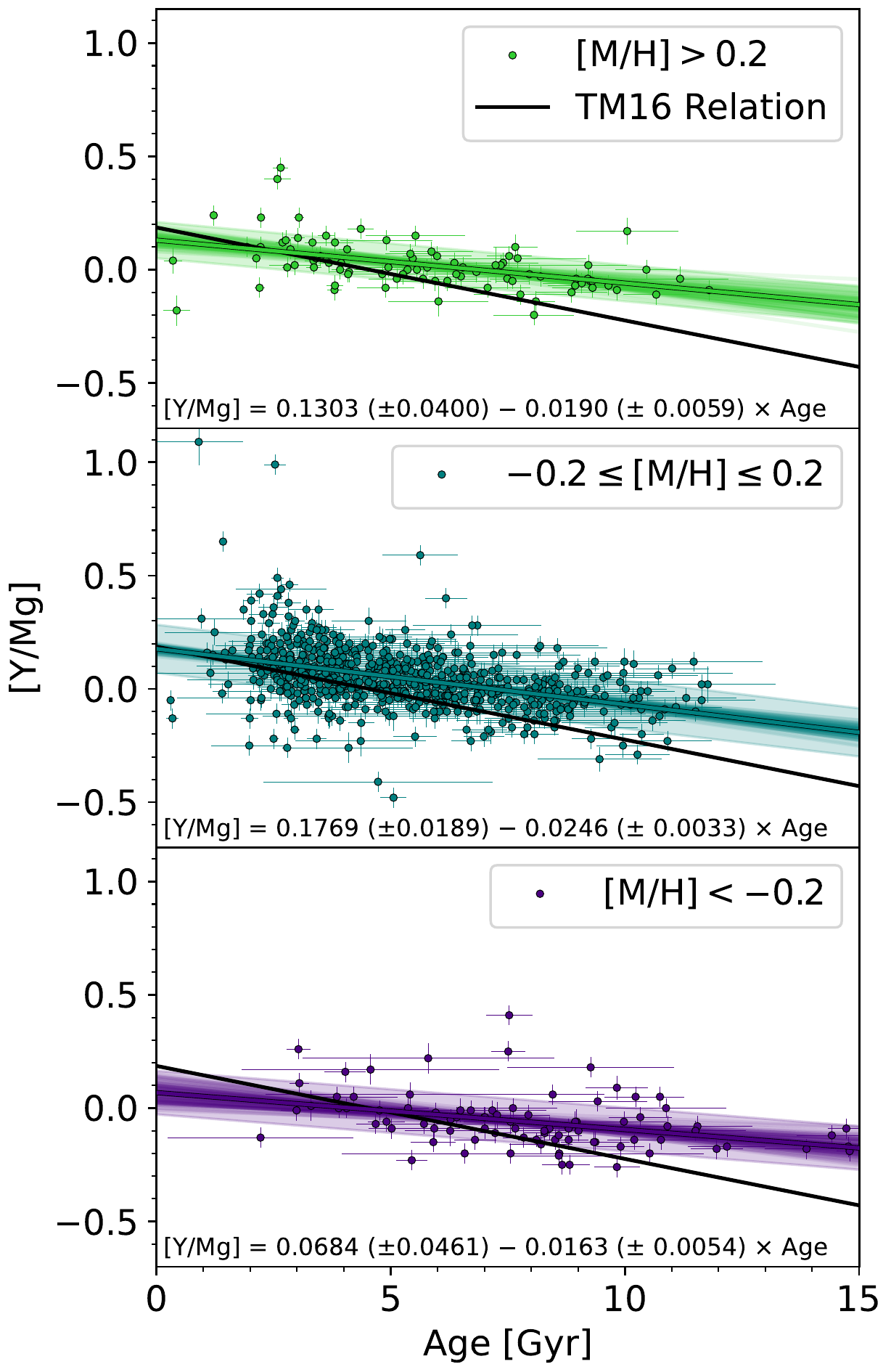}}
    \caption{[Y/Mg] versus stellar age for the reduced \ref{BF18} sample. The top, middle, and bottom panels show metallicities $>$ 0.2 dex, metallicities between --0.2 and 0.2 dex, and metallicities $<$ --0.2 dex, respectively. We plot the best fitting relations as solid lines and the MCMC realizations as translucent lines underneath. In black we plot the \ref{TM16} best fit relation.}
    \label{fig:ymgAgemet}
\end{figure}

\citet{Feltzing2017} demonstrated that the [Y/Mg] Clock appears to weaken for sub-solar metallicities of $\sim$\,--0.5\,dex. Therefore, we investigate the reduced \ref{BF18} sample for the presence of a metallicity-dependent slope in the [Y/Mg] Clock. Figure \ref{fig:ymgAgemet} shows the [Y/Mg]-Age relationship as a function of metallicity. We do not see any large differences in the age ranges of each metallicity sample. This is unsurprising given the rather flat stellar age-metallicity relation of the Galaxy \citep{GCS1,Haywood2013,Bergemann2014}.

The top, middle, and bottom panels display the super-solar, solar, and sub-solar metallicity stars in the reliable \ref{BF18} sample, respectively. In general, we do not find any significant trends as a function of metallicity. Only the solar metallicity sample exhibits a $>$1$\sigma$ difference from the low metallicity sample. We performed $F$-tests and found p-value significances all over 3$\sigma$, suggesting each sample is best described by a slope + intercept model over an intercept-only model. Summaries of the fit parameters can be found in Table \ref{tab:agefits}. Because only the low metallicity sample produces a marginally less-sensitive [Y/Mg]-Age relationship than the solar metallicity sample, our results are in general agreement with the \citet{Feltzing2017} conclusion that sub-solar metallicity stars produce weaker [Y/Mg]-Age relationships.

\section{[Y/Mg] and Stellar Rotation}
\label{sec:ymgrot}

\subsection{Rotation Sample}
\label{sec:ymgrotsamp}

\begin{figure}
    \centering
    \resizebox{\hsize}{!}{\includegraphics{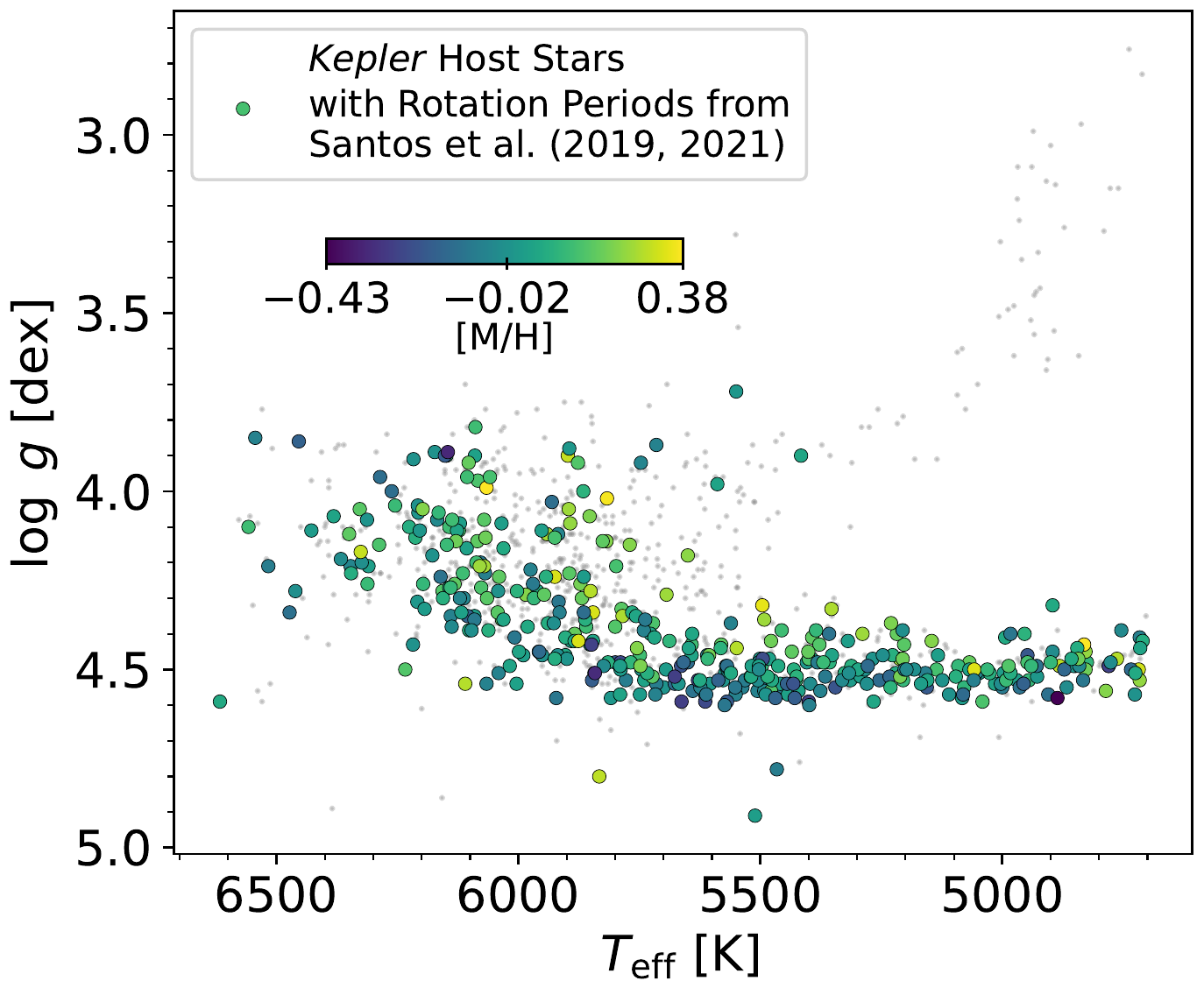}}
    \caption{Kiel diagram of \kep\ host stars with spectroscopic parameters and [Y/Mg] abundances from \ref{BF18}. The colored circles represent the 401 stars that have a measured rotation period in \citet{Santos2019} or \citet{Santos2021}, and are colored by their metallicity from \ref{BF18}. Plotted in grey are \kep\ host stars without measured rotation periods from \citet{Santos2019} or \citet{Santos2021}.}
    \label{fig:protsamp}
\end{figure}

\begin{figure}
    \centering
    \resizebox{\hsize}{!}{\includegraphics{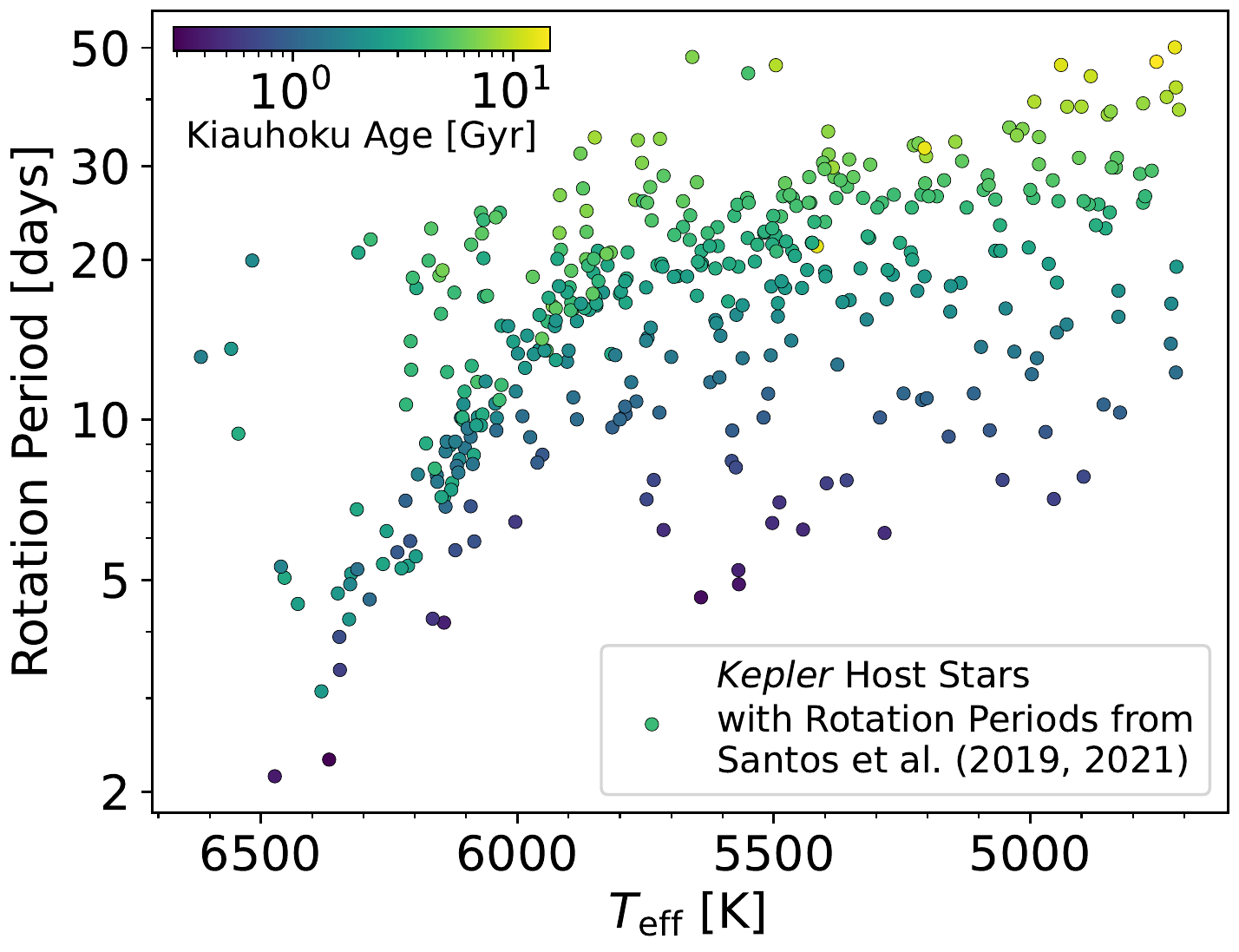}}
    \caption{Rotation period versus effective temperature for the sample 401 star sample highlighted in Figure \ref{fig:protsamp}. Stars are colored by their \texttt{kiauhoku} age on a logarithmic scale.}
    \label{fig:protteffkiau}
\end{figure}

Isochrone ages are not informative for lower main-sequence stars, as they do not evolve quickly enough. Therefore, we can try to use other, more sensitive age indicators for low-mass stars, such as stellar rotation periods \citep{Barnes2007,Mamajek2008,Soderblom2010,vansaders16,Curtis2019}, to compare directly against [Y/Mg].

First, we crossmatched the \ref{BF18} data with the rotation periods (\prot) provided by \citet{Santos2019} and \citet{Santos2021}, which supersedes the \citet{McQuillan2013} and \citet{McQuillan2014} catalogs. We plot this sample in Figure \ref{fig:protsamp}. The colored circles, which represent stars with rotation periods, span the full range in \teff, include a few subgiants and no giants, and range in metallicity from --0.43 to 0.38 dex. Unlike luminosity and \teff, rotation periods clearly evolve with time for stars cooler than 5400 K, even if their exact rotation period-age relations remain a subject of debate. Unfortunately, the \ref{TM16} sample has only seven stars with light-curve-constrained rotation periods \citep{Lorenzo2019}, so we do not analyze them here and instead focus on the \ref{BF18} sample.

We translated the measured rotation periods to rotation ages using \texttt{kiauhoku} \citep{Claytor2020} and the \ref{BF18} spectroscopic \teff, \logg, and metallicity. We used the fast launch YREC models of \cite{vansaders2013} with the weakened magnetic braking prescription of \cite{vansaders16} to produce rotation ages. Figure \ref{fig:protteffkiau} shows the same sample in \prot-\teff\ space, with the derived ages from \texttt{kiauhoku}. We observe a clustering of stars in a band going from 10 days at 6100 K up to 20 days at 5500 K with similar colors, representing ages common for stars in the \kep\ field and solar neighborhood. The stars that appear above that cluster are subgiants, as expected. The apparent pile-up at long periods was addressed in \citet{David2022} and appears to be the signature of weakened magnetic braking \citep{vansaders16}; we account for this weakened magnetic braking in the \texttt{kiauhoku} modeling.

\subsection{Comparing Rotation and Isochrone Ages}
\label{sec:agevsage}

\begin{figure}
    \centering
    \resizebox{\hsize}{!}{\includegraphics{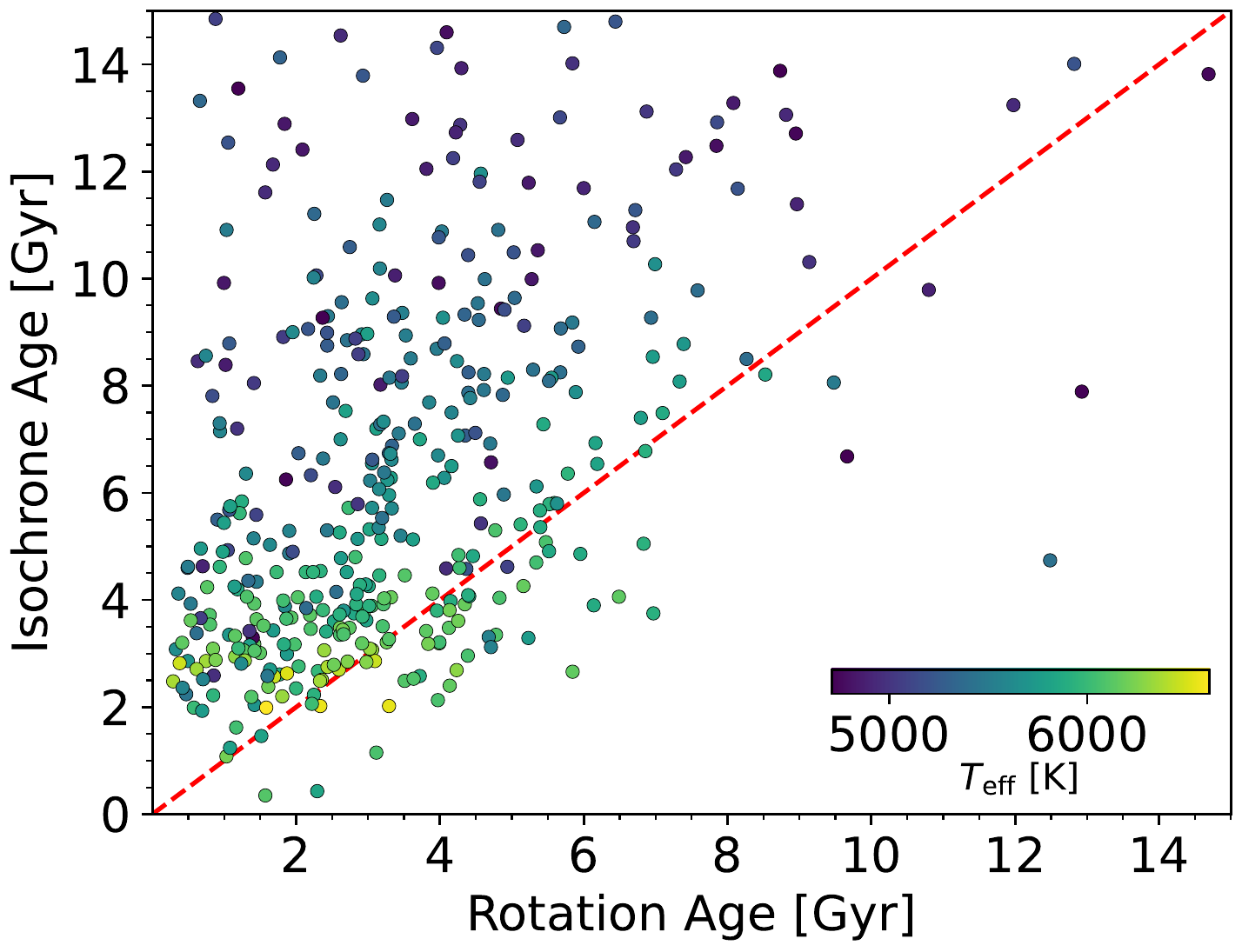}}
    \caption{Dartmouth isochrone ages from \ref{BF18} versus rotation ages from \texttt{kiauhoku}. Stars are colored by their \teff, and the 1:1 line is represented by the red, dashed line.}
    \label{fig:agevsage}
\end{figure}

In Figure \ref{fig:agevsage} we show a comparison of the \ref{BF18} isochrone ages and rotation ages from \texttt{kiauhoku}. We find that rotation-based ages are systematically younger than isochrone-based ages. The rotation age histogram peaks at $\approx$2.9 Gyr with only six stars older than 10 Gyr, while the \ref{BF18} isochrone age histogram peaks at $\approx$3.5 Gyr with 63 stars older than 10 Gyr. The solar neighborhood age distribution peaks around 3 Gyr and has a few stars older than 10 Gyr \citep[][, and references therein]{Lin2018}, which matches our rotation ages better than the \ref{BF18} isochrone ages. For stars at solar and cooler \teff, rotation is a more sensitive metric of age. Consequently, there is much better agreement for stars hotter than 5800 K, where isochrones are more sensitive to higher mass stars that evolve more quickly across the HR diagram. This may be part of the reason for why we find shallower relations above -- a cool, apparently old star with higher-than-expected [Y/Mg] may actually be a young star, but isochrone ages are too imprecise to tell the difference.

We caution that period-age relations remain a subject of debate. As the sample of stars with known ages and measured periods has increased, the picture of rotational evolution has become increasingly complex. \citet{vansaders16} found that stars past middle age appear to undergo dramatically reduced braking; \citet{Curtis2019} and \citet{curtis2020} highlighted that stars cooler than 5000 K appear to undergo a period of stalled spin-down at early to intermediate ages. Both phenomena affect the period-age relations. Our models account for the weakened braking in old stars; had we not included it we would have inferred even younger rotation-based ages. The stalled spin-down, which is most likely the effect of internal angular momentum transport \citep{denissenkov2010,spada2020} is not included, but should primarily impact stars cooler than the bulk of our sample.

\subsection{[Y/Mg]-Rotation Age Results}
\label{sec:ymgrotclock}

\begin{figure*}
    \centering
    \resizebox{\hsize}{!}{\includegraphics{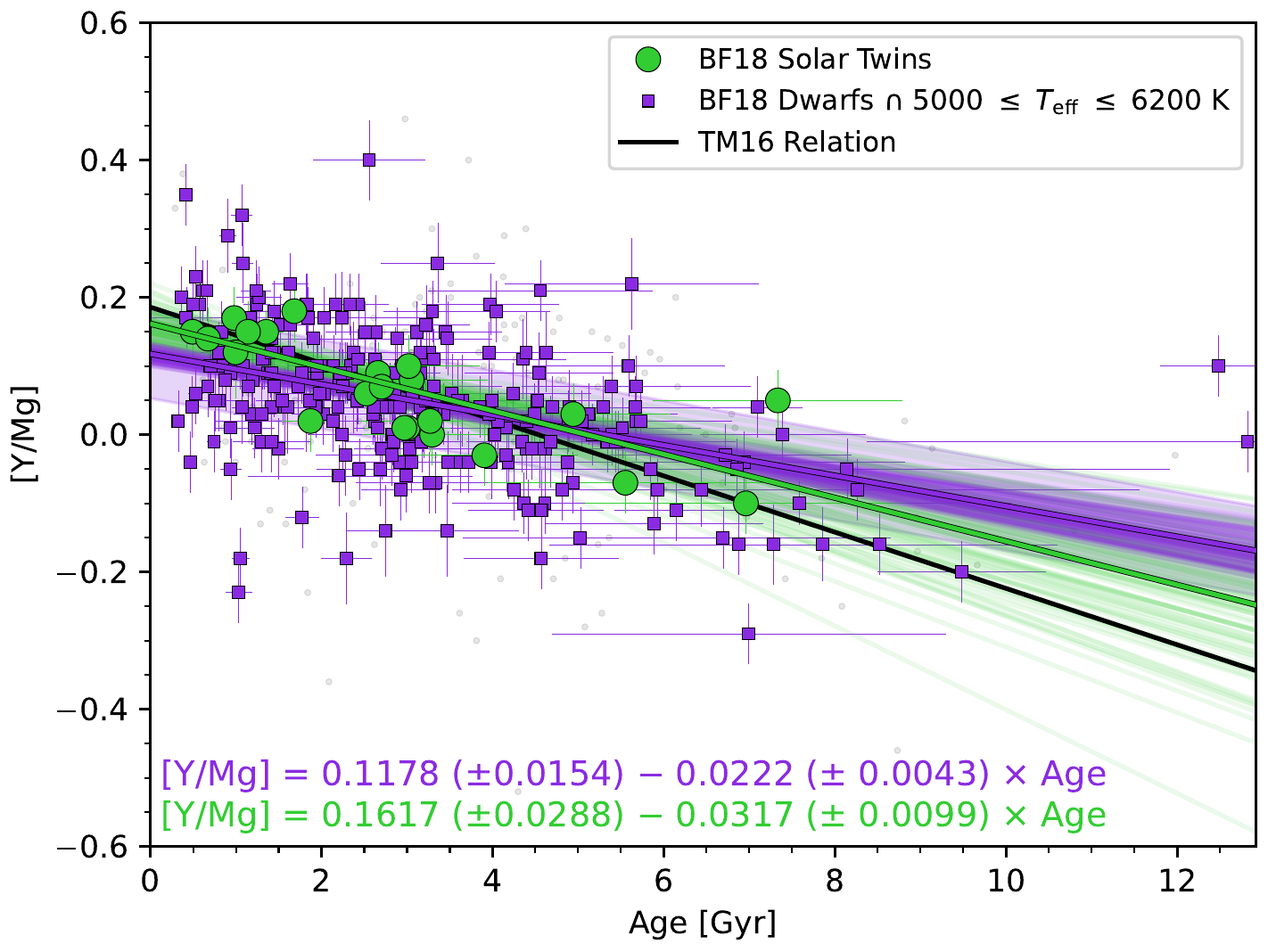}}
    \caption{[Y/Mg] versus rotation age for 22 \kep\ host star solar twins \citep[green circles; rotation periods from][and ages from \texttt{kiauhoku}]{Santos2019,Santos2021} and 273 \kep\ host stars with 5000 $\leq$ \teff\ $\leq$ 6200 K and $R_\star$ $<$ 1.4 $\mathrm{R}_\odot$ (purple squares). We plot the color-matched best fit relations, in addition to the \ref{TM16} best-fit relation in black and the full rotation sample as grey dots.}
    \label{fig:ymgrotage}
\end{figure*}

We performed MCMC analyses and bootstrap simulations as described in \S\ref{sec:ymgage}. Table \ref{tab:protfits} includes our best-fit relations for [Y/Mg] versus rotation ages, from the full 401 star sample described by Figures \ref{fig:protsamp} and \ref{fig:protteffkiau}, to solar twins and analogs. In particular, we choose a sample of stars between 5000 and 6200 K because stars cooler than 5000 K may experience core-envelope decoupling Gyrs into their evolution, while those hotter than 6200 K should not experience spin down while on the main sequence due to the lack of a convective envelope. We have also selected a sample of dwarfs (defined as stars with \texttt{kiauhoku}-derived radii below 1.4$\mathrm{R_\odot}$) between 5000 and 6200 K to avoid the complications of stellar evolution on rotation periods \citep{vansaders2013}.

Figure \ref{fig:ymgrotage} shows the 22 \ref{BF18} solar twins and 273 \ref{BF18} FGK dwarfs ($R_\star$ $<$ 1.4 $\mathrm{R}_\odot$, 5000 $\leq$ \teff\ $\leq$ 6200 K) with rotation ages from \texttt{kiauhoku}. For the solar twin and broader samples we compute slopes of $m$ = --0.0317 $\pm$ 0.0099 and --0.0222 $\pm$ 0.0043 dex/Gyr, intrinsic scatters of $c$ $<$ 0.001 and $c$ = 0.065 dex, age scatters of 2.3 and 4.6 Gyr, and $F$-test significances of 4.2 and 8.7$\sigma$, respectively. We have included the full results in rows three and five of Table \ref{tab:protfits}.

We also analyze subsamples of the rotating FGK dwarf sample, splitting the data into bins of \teff\ and metallicity, similar to our isochrone age analysis above. As a function of \teff, we find that the solar \teff\ stars produce the most sensitive relation, with small intrinsic scatter ($c$ = 0.039 dex), age scatter ($\sigma_{\mathrm{Age}}$ = 3.2 Gyr), and a 5.7$\sigma$ significant $F$-test p-value. The high- and low-\teff\ bins produce statistically consistent slopes, although the high-\teff\ stars produce a much more uncertain relation that does not prefer the slope + intercept model over the intercept-only model at 3$\sigma$ significance.

As a function of metallicity, we find statistically similar slopes for the super-solar, solar, and sub-solar metallicity bins. In particular, the sub-solar metallicity sample's slope is $<$1$\sigma$ significant and its $F$-test favors a slope + intercept fit at only 0.8$\sigma$, suggesting there is no significant trend in [Y/Mg] as a function of age. The super-solar metallicity sample has the most sensitive relation and the lowest corresponding age uncertainty (3.8 Gyr), but given the lack of a clear trend as a function of metallicity, it is hard to conclude anything confidently.

Overall, we find that the most sensitive and tightest [Y/Mg]-Rotation Age relations occur for solar twins and analogs. For solar twins, we find a relation that is mostly consistent with the \ref{TM16} isochrone age-based results, and similarly suggests that [Y/Mg] is a reliable age indicator for solar twins. However, as we find above, the [Y/Mg]-Age relationship weakens for non-solar twin dwarfs and subgiant stars. In addition, we find no metallicity trend, which matches the above isochrone metallicity results.

\section{Discussion \& Conclusion}

We confirm that [Y/Mg] works best as a clock for solar twins, but the relationship we find is not as sensitive or as significant as those reported by \citet{Nissen2015}, \ref{TM16}, and \citet{Spina2016}. We also find that the [Y/Mg]-Age relation weakens for non-solar type stars in \teff, \logg, and metallicity, in agreement with \citet{Feltzing2017}. Unfortunately, the \ref{BF18} sample is composed solely of \kep\ host stars, and their location in the solar neighborhood \citep[most within 1 kpc,][]{brewer15} prevents us from directly comparing our results to \citet{Anders2018} and \citet{Titarenko2019} in evaluating Galactic location-dependent [Y/Mg]-Age relationships.

In addition, the slope for \ref{BF18} sub-solar \logg\ stars ($m$ = --0.0309 $\pm$ 0.0037 dex/Gyr) is statistically consistent with the \ref{BF18} solar twins slope ($m$ = --0.0370 $\pm$ 0.0071 dex/Gyr), in agreement with \citet{Slumstrup2017}, which found a similar [Y/Mg]-Age trend for helium-core-burning giant stars in open clusters as the solar twins in \ref{TM16}. We do note that the \ref{BF18} isochrone fitting was not specifically tuned for giant stars, where different physical ingredients in the models become important when trying to determine an isochrone age \citep{Tayar2017,Choi2018b,Tayar2022}, preventing a direct comparison to the results of \replaced{\citet{Slumstrup2017}}{\citet{Slumstrup2017,Casamiquela2021}}. We also are wary of the isochrone method's varying sensitivity over the range of \teff\ and \logg\ investigated here and its potential to bias our best-fit relations. In general, we show that the [Y/Mg]-Age relation is useful but of varying utility depending on the stellar sample of interest.

\begin{figure*}[h]
    \centering
    \resizebox{\hsize}{!}{\includegraphics{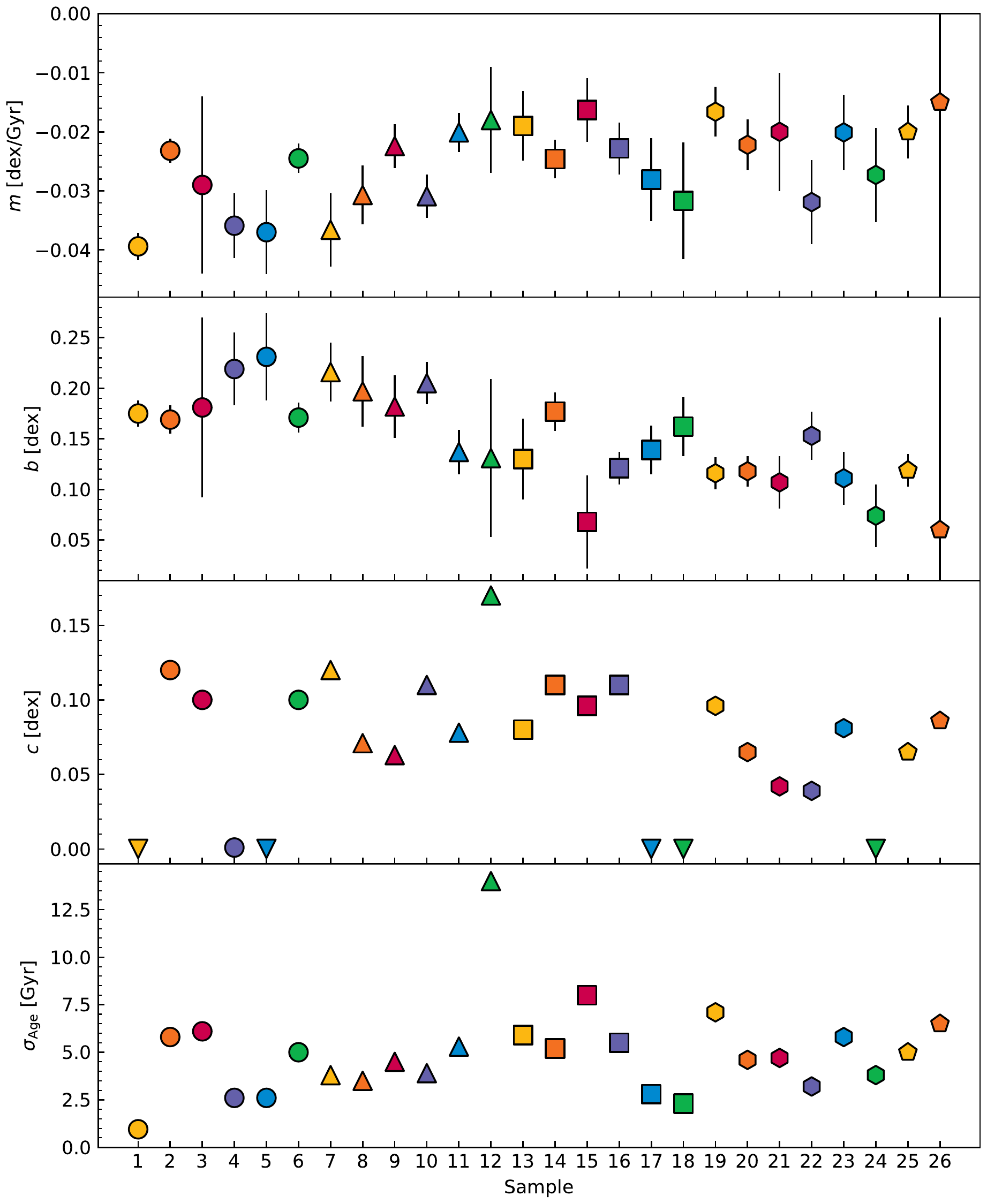}}
    \caption{Summary of the best-fit relations detailed in Tables \ref{tab:agefits} and \ref{tab:protfits}. Samples are numbered according to the numerical value in the Sample column of each table. Panels from top to bottom display the slope (and uncertainty), intercept (and uncertainty), intrinsic scatter, and age uncertainty of each best-fit relation, respectively. In the intrinsic scatter panel, upper limits ($c$ $<$ 0.001 dex) are displayed as downward triangles.}
    \label{fig:fitsummary}
\end{figure*}

We present the first comparison of [Y/Mg] and rotation ages and find that the behavior of [Y/Mg] with rotation age is consistent with isochrone age comparisons, where the [Y/Mg] clock performs best for solar twins \citep{Nissen2015,Maia2016,Spina2016,Feltzing2017}. This holds true even though rotation and isochrone ages differ significantly, especially for sub-solar mass stars. Much like with isochrone ages, [Y/Mg] is typically not as sensitive or precise as a rotation age diagnostic for stars that are not solar twins. We also find significant differences between isochrone and rotation age estimates. We suggest that (1) [Y/Mg]-Isochrone age relations should be used for subgiants and (2) [Y/Mg]-Rotation age relations should be used for sub-solar mass dwarfs. For solar-type dwarfs, both rotation and isochrone [Y/Mg]-Age relations appear to perform similarly. Tables \ref{tab:agefits} and \ref{tab:protfits} and Figure \ref{fig:fitsummary} summarize our best-fit relations.

Finally, we note that binaries are unlikely to have a significant impact on the analysis done in this paper, as \citet{Furlan2017} determined that $\approx$10\% of \kep\ hosts have AO-detected binary companions within 1'', which is larger than the slit-width used in the \ref{BF18} spectra \citep[0.86'',][]{Petigura2017}. In addition, only a subset of these stars will affect the \ref{BF18} spectroscopic analysis, as (1) double-lined spectroscopic binaries are not included in \ref{BF18} \citep[removed by][]{Petigura2017} and (2) only systems with moderate mass ratios should produce suspect stellar properties and abundances. Binaries impact the rotation sample even less than the isochrone sample, as \citet{Santos2019,Santos2021} flag only seven stars as Classical Pulsator/Close-in Binary candidates and the rest do not have a binary flag in either \citet{Berger2018c} or \citet{Simonian2019}. For wide binaries, where we do not expect there to be significant impact on stellar parameters or measured abundances, \citet{Espinoza-Rojas2021} shows that even coeval, chemically inhomogeneous binaries can display consistent chemical clocks.

We summarize our conclusions as follows:
\begin{enumerate}
    \item We find that [Y/Mg]-Age relations for solar twins and analogs are the most sensitive ($m$ $\lesssim$ --0.03 dex/Gyr) and tightest ($c$ $<$ 0.001 dex, $\sigma_{\mathrm{Age}}$ $<$ 2.8 Gyr) among the FGK stars analyzed in this paper. We find that the \ref{BF18} \kep\ solar twins produce a [Y/Mg]-Age slope that is statistically consistent with the \ref{TM16} relation, albeit with an intercept offset that would produce ages different by $\gtrsim$1.5 Gyr given a measured [Y/Mg]. We note that any differences between our relation and those in the literature could be due to non-solar twin contaminants in our sample or systematics in the abundance or age determinations. We also compared [Y/Mg] to rotation age for the first time and produce similar findings.
    
    \item We do not find any significant trends in the [Y/Mg] Clock as a function of \teff, \logg, or metallicity. In general, we find that non-solar FGK-type samples produce \replaced{less sensitive}{shallower} ($m$ $\gtrsim$ -- 0.02 dex/Gyr) [Y/Mg]-Age relations with greater intrinsic scatter ($c$ $>$ 0.04 dex) and age scatter ($\sigma_{\mathrm{Age}}$ $>$ 3.2 Gyr) than the solar twin and analog samples. However, many of these relations remain statistically significant and may be useful for future related work.
    
    \item We compare isochrone and rotation ages for a subsample of \ref{BF18} stars with rotation periods and find significant differences between the isochrone and rotation age estimates. In particular, the rotation ages are systematically younger and match the age of the solar neighborhood better than isochrone ages. We suggest that [Y/Mg]-Isochrone age relations should be used for subgiants, [Y/Mg]-Rotation age relations should be used for sub-solar mass dwarfs, and both rotation and isochrone ages appear to perform similarly for solar-type dwarfs. Finally, we compared our solar twin [Y/Mg]-Isochrone age relation to the literature relations \citep[Table 6 of][]{Delgado2019} and found a corresponding systematic age uncertainty of $\approx$1.5 Gyr, which we suggest to add in quadrature to any age derived from a [Y/Mg]-Age relation.
\end{enumerate}

While the [Y/Mg] Clock is most useful for solar twins, it can still be applied to more diverse FGK stars. It also does not preclude the use of an ensemble of age indicators to more robustly constrain the ages of field stars. We look forward to future investigations leveraging the now-diverse set of age indicators for $Kepler$ and soon $TESS$ planet host stars.

\software{\texttt{emcee} \citep{Foreman-Mackey2013}, \texttt{kiauhoku} \citep{Claytor2020}, \texttt{matplotlib} \citep{Matplotlib}, \texttt{numpy} \citep{numpy}, \texttt{pandas} \citep{Pandas}, \texttt{scipy} \citep{Scipy}, \texttt{TOPCAT} \citep{topcat}}

\begin{acknowledgments}
We thank the referee for the useful comments that improved the paper. We also thank Eugene Magnier, Christoph Baranec, Aleezah Ali, Tom Barclay, Vanshree Bhalotia, Casey Brinkman, Ashley Chontos, Marc Hon, Maryum Sayeed, Nicholas Saunders, Aldo Sepulveda, Jamie Tayar, Lauren Weiss, and Jingwen Zhang for helpful discussions in addition to feedback on the figures. A portion of T.A.B.’s research was supported by an appointment to the NASA Postdoctoral Program at the NASA Goddard Space Flight Center, administered by Universities Space Research Association and Oak Ridge Associated Universities under contract with NASA. T.A.B. and D.H. acknowledge support by a NASA FINESST award (80NSSC19K1424). D.H. also acknowledges support from the Alfred P. Sloan Foundation and NASA grant 80NSSC19K0597. E.G. acknowledges support from NSF award AST-187215.
\end{acknowledgments}

\begin{deluxetable*}{lccccccr}
\tabletypesize{\scriptsize}
\tablenum{1}
\tablecolumns{8}
\tablecaption{[Y/Mg]--Isochrone Age Best-Fit Relations}
\tablehead{
\colhead{Sample} & \colhead{Slope ($m$)} & \colhead{$\sigma_m$} & \colhead{Intercept ($b$)} & \colhead{$\sigma_b$} & Intrinsic Scatter ($c$) & $\sigma_{\mathrm{Age}}$ (Gyr) & $F$-Test ($\sigma$)}
\def\arraystretch{1.0}
\startdata
1. \ref{TM16} & --0.0394 & 0.0023 & 0.175 & 0.013 & $<$0.001 & 0.95 & 12.4 \\
2. \ref{BF18} & --0.0232 & 0.0020 & 0.169 & 0.014 & 0.12 & 5.8 & 18.6 \\
3. Asteroseismic & --0.029 & 0.015 & 0.181 & 0.089 & 0.10 & 6.1 & 2.9 \\
4. \ref{BF18} Solar Analogs & --0.0359 & 0.0055 & 0.219 & 0.036 & 0.001 & 2.6 & 6.1 \\
5. \ref{BF18} Solar Twins & --0.0370 & 0.0071 & 0.231 & 0.043 & $<$0.001 & 2.6 & 4.2 \\
6. Reliable \ref{BF18} Sample (\teff\ $>$ 5400 K) & --0.0245 & 0.0025 & 0.171 & 0.015 & 0.10 & 5.0 & 14.5 \\
7. Reliable \ref{BF18} Sample $\cap$ \teff\ $>$ 5872 K & --0.0366 & 0.0062 & 0.216 & 0.029 & 0.12 & 3.8 & 8.9 \\
8. Reliable \ref{BF18} Sample $\cap$ 5672 $\leq$ \teff\ $\leq$ 5872 K & --0.0307 & 0.0050 & 0.197 & 0.035 & 0.071 & 3.5 & 8.6 \\
9. Reliable \ref{BF18} Sample $\cap$ \teff\ $<$ 5672 K & --0.0224 & 0.0037 & 0.182 & 0.031 & 0.063 & 4.5 & 7.4 \\
10. Reliable \ref{BF18} Sample $\cap$ \logg\ $<$ 4.34 & --0.0309 & 0.0037 & 0.205 & 0.021 & 0.11 & 3.9 & 11.8 \\
11. Reliable \ref{BF18} Sample $\cap$ 4.34 $\leq$ \logg\ $\leq$ 4.54 & --0.0201 & 0.0033 & 0.137 & 0.022 & 0.078 & 5.3 & 7.9 \\
12. Reliable \ref{BF18} Sample $\cap$ \logg\ $>$ 4.54 & --0.0180 & 0.0090 & 0.131 & 0.078 & 0.17 & 14 & 3.0 \\
13. Reliable \ref{BF18} Sample $\cap$ [M/H] $>$ 0.2 & --0.0190 & 0.0059 & 0.130 & 0.040 & 0.080 & 5.9 & 5.2 \\
14. Reliable \ref{BF18} Sample $\cap$ --0.2 $\leq$ [M/H] $\leq$ 0.2 & --0.0246 & 0.0033 & 0.177 & 0.019 & 0.11 & 5.2 & 11.4 \\
15. Reliable \ref{BF18} Sample $\cap$ [M/H] $<$ --0.2 & --0.0163 & 0.0054 & 0.068 & 0.046 & 0.096 & 8.0 & 3.5
\enddata
\tablecomments{Best-fit relations computed for the various [Y/Mg]-Isochrone Age comparisons detailed in this paper. All equations are of the form [Y/Mg] = $m$ $\times$ Age + $b$, and 1$\sigma$ uncertainties are quoted for each parameter. We fit for intrinsic scatter by adding the term $c^2 \cos^2{\left(\arctan{\left(m\right)}\right)}$ to the variance in our MCMC analysis (see \S \ref{sec:ymgage}), and report $\sigma_{\mathrm{Age}}$, which is the corresponding scatter in age in units of Gyr about the best-fit relation. We also include our $F$-test results in the final column, indicating the corresponding significance of the $p$-value in units of $\sigma$ at which the data prefer two-parameter fits (slope plus intercept) over one-parameter fits (intercept only). We plot the summary statistics for this table's rows 1--15 in columns 1--15 of Figure \ref{fig:fitsummary}, respectively.}\label{tab:agefits}
\end{deluxetable*}

\begin{deluxetable*}{lccccccr}
\tabletypesize{\scriptsize}
\tablenum{2}
\tablecolumns{8}
\tablecaption{[Y/Mg]--Rotation Age Best-Fit Relations}
\tablehead{
\colhead{Sample} & \colhead{Slope ($m$)} & \colhead{$\sigma_m$} & \colhead{Intercept ($b$)} & \colhead{$\sigma_b$} & Intrinsic Scatter ($c$) & $\sigma_{\mathrm{Age}}$ (Gyr) & $F$-Test ($\sigma$)}
\def\arraystretch{1.0}
\startdata
16. \ref{BF18} & --0.0228 & 0.0044 & 0.121 & 0.016 & 0.11 & 5.5 & 8.6 \\
17. \ref{BF18} Solar Analogs & --0.0281 & 0.0070 & 0.139 & 0.024 & $<$0.001 & 2.8 & 4.7 \\
18. \ref{BF18} Solar Twins & --0.0317 & 0.0099 & 0.162 & 0.029 & $<$0.001 & 2.3 & 4.2 \\
19. \ref{BF18} 5000 $\leq$ \teff\ $\leq$ 6200 K & --0.0166 & 0.0042 & 0.116 & 0.016 & 0.096 & 7.1 & 6.2 \\
20. \ref{BF18} Dwarfs $\cap$ 5000 $\leq$ \teff\ $\leq$ 6200 K & --0.0222 & 0.0043 & 0.118 & 0.015 & 0.065 & 4.6 & 8.7 \\
21. \ref{BF18} Dwarfs $\cap$ 5900 $\leq$ \teff\ $\leq$ 6200 K & --0.020 & 0.010 & 0.107 & 0.026 & 0.042 & 4.7 & 2.5 \\
22. \ref{BF18} Dwarfs $\cap$ 5600 $\leq$ \teff\ $\leq$ 5900 K & --0.0319 & 0.0071 & 0.153 & 0.024 & 0.039 & 3.2 & 5.7 \\
23. \ref{BF18} Dwarfs $\cap$ 5000 $\leq$ \teff\ $\leq$ 5600 K & --0.0201 & 0.0064 & 0.111 & 0.026 & 0.081 & 5.8 & 5.7 \\
24. \ref{BF18} Dwarfs $\cap$ 5000 $\leq$ \teff\ $\leq$ 6200 K $\cap$ [M/H] $>$ 0.2 & --0.0273 & 0.0080 & 0.074 & 0.031 & $<$0.001 & 3.8 & 4.3 \\
25. \ref{BF18} Dwarfs $\cap$ 5000 $\leq$ \teff\ $\leq$ 6200 K $\cap$ --0.2 $\leq$ [M/H] $\leq$ 0.2 & --0.0200 & 0.0045 & 0.119 & 0.016 & 0.065 & 5.0 & 7.4 \\
26. \ref{BF18} Dwarfs $\cap$ 5000 $\leq$ \teff\ $\leq$ 6200 K $\cap$ [M/H] $<$ --0.2 & --0.015 & 0.046 & 0.06 & 0.21 & 0.086 & 6.5 & 0.8 \\
\enddata
\tablecomments{Best-fit relations computed for the various [Y/Mg]-Rotation Age comparisons detailed in Section \ref{sec:ymgrot}. All equations are of the form [Y/Mg] = $m$ $\times$ Age$_{P_{\mathrm{rot}}}$ + $b$, and 1$\sigma$ uncertainties are quoted for each parameter. We fit for intrinsic scatter by adding the term $c^2 \cos^2{\left(\arctan{\left(m\right)}\right)}$ to the variance in our MCMC analysis, and report $\sigma_{\mathrm{Age}}$, which is the corresponding scatter in age in units of Gyr about the best-fit relation. We also include our $F$-test results in the final column, indicating the corresponding significance of the $p$-value in units of $\sigma$ at which the data prefer two-parameter fits (slope plus intercept) over one-parameter fits (intercept only). We plot the summary statistics for this table's rows 16--26 in columns 16--26 of Figure \ref{fig:fitsummary}, respectively.} \label{tab:protfits}
\end{deluxetable*}

\bibliography{references}{}
\bibliographystyle{aasjournal}


\end{document}